\documentclass{elsart}

\usepackage{graphicx}
\usepackage{amssymb}
\usepackage{epsfig}

\def\Journal#1#2#3#4{{#1} {#2} (#3) #4}

\def\NPA{Nucl. Phys. A}
\def\PLB{Phys. Lett.  B}

\def\PRC{Phys. Rev. C}

\def\be{\begin{equation}}
\def\ee{\end{equation}}

\newcommand{\ud}{\mathrm{d}}

\begin{document}
\begin{frontmatter}

\title{Statistical hadronization of heavy quarks in ultra-relativistic
  nucleus-nucleus collisions}

\author[gsi]{A.~Andronic}, %\thanksref{info}},
\author[gsi,tud]{P.~Braun-Munzinger},
\author[wro]{K.~Redlich},
\author[hei]{J.~Stachel}

\address[gsi]{Gesellschaft f\"ur Schwerionenforschung, GSI, 
D-64291 Darmstadt, Germany}
\address[tud]{Technical University Darmstadt, D-64289 Darmstadt, Germany}
\address[wro]{Institute of Theoretical Physics, University of Wroc\l aw,
PL-50204 Wroc\l aw, Poland}
\address[hei]{Physikalisches Institut der Universit\"at Heidelberg,
D-69120 Heidelberg, Germany}

\begin{abstract}
  We present new results on the statistical hadronization of heavy quarks at
  SPS, RHIC and LHC energies.  Several new aspects are considered, among them
  a separation of the collision geometry into a ``core'' and a ``corona'' part
  and an estimate of the annihilation rate of charm quark in a hot plasma,
  together with a critical assessment of its influence on the results.  For
  RHIC energies we investigate the centrality dependence of $J/\psi$
  production focusing on the model results for different values of the charm
  production cross section, including its theoretical and experimental
  uncertainty. We also study, within this model, the rapidity dependence of
  the $J/\psi$ yield. Recent RHIC data from the PHENIX experiment are well
  reproduced.  At LHC energy, we update our model predictions for the
  centrality dependence of the $J/\psi$ yield and investigate as well the
  rapidity dependence.  We also discuss the transverse momentum distributions
  of $J/\psi$ mesons expected from the model and provide predictions for a
  range of values of the expansion velocity at chemical freeze-out.  Finally,
  we extend the model to predict $\Upsilon$ yields in Pb+Pb collisions at 
  LHC energy.
\end{abstract}

\vspace{2mm}
%PACS: {25.75.Ld;25.70.Pq}

\end{frontmatter}

\section{Introduction}

The idea of statistical hadronization of charm quarks in nucleus-nucleus
collisions \cite{pbm1} has led to a series of investigations of $J/\psi$
production based on this approach \cite{gor1,gra1,kos,aa1,bra}.  Initial
interest focussed on the available SPS data for $J/\psi$ production in Pb-Pb
collisions, but the trends for RHIC and LHC were also investigated.  An
independent approach, based on a kinetic model \cite{the1,the2,the3} has been
developed in parallel.  Recently the statistical hadronization model was
extended to the production of $\Upsilon$ mesons \cite{gra2} and multiply
heavy-flavored hadrons \cite{bec}.

We have shown earlier \cite{pbm1,aa1} that the $J/\psi$ data at SPS
energy can be described within the statistical approach, but only when
assuming that the charm production cross section is enhanced by about
a factor of 3 beyond the perturbative QCD (pQCD) predictions.  For
RHIC energy, the predictions of our model were shown \cite{aa1} to be
in good agreement with the early PHENIX $J/\psi$ data \cite{phe1}
as well as with the relative yields of open charm mesons measured by
STAR \cite{tai}.

We focus in this paper on the production of charmonia in
nucleus-nucleus collisions at RHIC (Au-Au, $\sqrt{s_{NN}}$=200 GeV)
and LHC (Pb-Pb, $\sqrt{s_{NN}}$=5.5 TeV) energies within the
framework of the statistical hadronization model, SHM
\cite{pbm1,aa1}. Particular emphasis is placed on the rapidity and
transverse momentum dependence of $J/\psi$ and, for LHC energy, also
$\Upsilon$  production.  
Section 2 contains a brief summary of the main ingredients of the model along
with a description of recent updates concerning the chemical freeze-out
parameters, the spectrum of open charm mesons and baryons, and the
values of charm production cross section.  Furthermore, for a more
realistic description of the centrality dependence in nucleus-nucleus 
collisions, we separate the collision geometry into a 'core' and a 'corona' 
part.  

In Section 3, we provide an estimate of the magnitude of
$c\bar{c}$ annihilation in the expanding quark-gluon plasma (QGP).

In Section 4, we use the model to predict phase space
distributions of quarkonia produced in nucleus-nucleus collisions.
We compare the results with SPS data and with data available at RHIC 
\cite{per} and provide predictions for the LHC energy, where data 
\cite{cro} are expected in about two years.  In particular the situation
at SPS energy is significantly modified when taking into account the 
realistic reaction geometry. 
We extend the model for the calculation of $\Upsilon$ yields in Pb+Pb 
collisions at LHC energy.  
In this case, despite the similarity in production rates
of bottom at LHC with charm at RHIC \cite{cro2}, the applicability  of our
model may be questionable, as the thermalization of the bottom quarks 
may be debatable. We discuss below the main sources of uncertainty in 
these calculations.

\section{The model and its inputs}

The statistical hadronization model (SHM) \cite{pbm1,aa1} assumes that all
heavy quarks (charm and bottom) are produced in primary hard collisions and
that their total number stays constant until hadronization.  The validity of
this latter point is investigated in the next section.  Another important
factor is thermal (but not chemical) equilibration in the QGP, at least near
the critical temperature, $T_c$. The quarkonia are then all produced
(non-perturbatively) when $T_c$ is reached and the system hadronizes.  Our
picture is different from the one of color screening \cite{satz} (see
\cite{satz2} for a more recent account), which assumes that formation of
quarkonia takes place before a thermalized QGP is established but is
suppressed in the QGP when the temperature reaches a certain threshold, which
is species dependent.  Recent results from solving Quantum Chromodynamics
(QCD) on the lattice  indicate indeed that $J/\psi$ mesons could survive
in the QGP up to 1.6$T_c$ \cite{asa}, implying, e.g. little melting at SPS
energies.  Further investigations indicate that $\Upsilon$ mesons may survive
up to temperatures of at least 2.3$T_c$ \cite{pet}.  

In the SHM, we assume that (i) no quarkonia production takes place before the
formation time $\tau_0$ of the QGP or (ii) that all quarkonia formed before
$\tau_0$ are melted in the initial hot QGP phase. Under this scenario the
possible existence of bound quarkonia states in the QGP is not important,
since the estimates made in the next section demonstrate that, unless there is
a resonance-like enhancement of the cross section, formation of quarkonia from
uncorrelated $c$ and $\bar c$ quarks in the expanding plasma is very unlikely
even if such states exist as bound states.

Thermal equilibration of charm quarks in a QGP \footnote{The cross
  sections for production of charmed quarks  are much too small to allow
  for their chemical equilibration  in a QGP at reasonable
  ($T<$1 GeV) temperatures, as is demonstrated in \cite{redlich_pbm}.} 
is currently investigated vigorously both experimentally and theoretically.
In particular, elliptic flow of charm quarks \cite{bat} has been demonstrated
within a coalescence approach \cite{v2x,v2y} to be a very sensitive probe of
thermalization.  Large elliptic flow is obtained in hydrodynamic simulations
based on the diffusion coefficient of charm quarks in a QGP \cite{v2z}.
Elliptic flow of $J/\psi$ is smaller in recent hydrodynamic calculations
\cite{nu} than in coalescence models \cite{bz}.  Remarkably, the experimental
results on the elliptic flow of electrons from charm mesons ("non-photonic"
electrons) \cite{v2b,v2b2} do indicate that charm quarks indeed thermalize to a
significant degree.  
In addition, a large energy loss of charm quarks in the QGP \cite{djo} has 
been recently found by experiments \cite{v2b2,raa1,raa2}.  The experimental 
results of PHENIX \cite{raa1} indicate in fact a larger energy loss than 
theoretically expected \cite{djo1}. The mechanism for charm thermalization 
is not well understood.  A model based on resonant rescattering of heavy 
quarks in the QGP \cite{vhr} reproduces quite well the experimental results.
It is worth noticing that in this model the thermalization time for heavy 
quarks is substantially reduced compared to the case of perturbative 
interactions \cite{vhr}.  In transport models the measured elliptic flow 
can be reproduced only for a charm quark scattering cross section that is 
much larger than the cross sections computed within perturbative QCD
\cite{zha,molnar}.  In any case, we note that the SHM can  be applied even 
if charm thermalization is reached only close to $T_c$.

In the following we briefly outline the calculation steps in our model
\cite{pbm1,aa1}.  The heavy quark (charm or bottom) balance equation 
\cite{pbm1}, which has to include canonical suppression factors whenever 
the number of charm or bottom pairs is not much larger than 1, is used 
to determine a fugacity  factor $g_c$ via:
\begin{equation}
N_{c\bar{c}}^{dir}=\frac{1}{2}g_c N_{oc}^{th}
\frac{I_1(g_cN_{oc}^{th})}{I_0(g_cN_{oc}^{th})} + g_c^2N_{c\bar c}^{th}.
\label{aa:eq1}
\end{equation}
Here $N_{c\bar{c}}^{dir}$  is the number of directly produced $c\bar{c}$ 
pairs and  $I_n$ are  modified Bessel functions. In the fireball of volume 
$V$ the total number of open ($N_{oc}^{th}=n_{oc}^{th}V$) and hidden 
($N_{c\bar c}^{th}=n_{c\bar c}^{th}V$) charm hadrons is computed from 
their grand-canonical densities $n_{oc}^{th}$ and $n_{c\bar c}^{th}$, 
respectively. 
The densities of different particle species in the grand canonical ensemble 
are calculated following the statistical model \cite{pbm0,pbm0a,aa2}.
The balance equation (\ref{aa:eq1}) defines the fugacity parameter $g_c$ 
that accounts for deviations of heavy quark multiplicity from the value 
that is expected in complete chemical equilibrium. Since in the SHM the 
charm and bottom quark yields are completey determined by their production 
in initial (hard) collisions the fugacity factor can be very large, 
especially for bottom production. 
The yield of charmonia of type $j$ is obtained as: $N_j=g_c^2 N_j^{th}$.

\begin{figure}[htb]
\begin{tabular}{lr}\begin{minipage}{.49\textwidth}
\vspace{-1cm}
\hspace{-0.5cm}\includegraphics[width=1.15\textwidth]{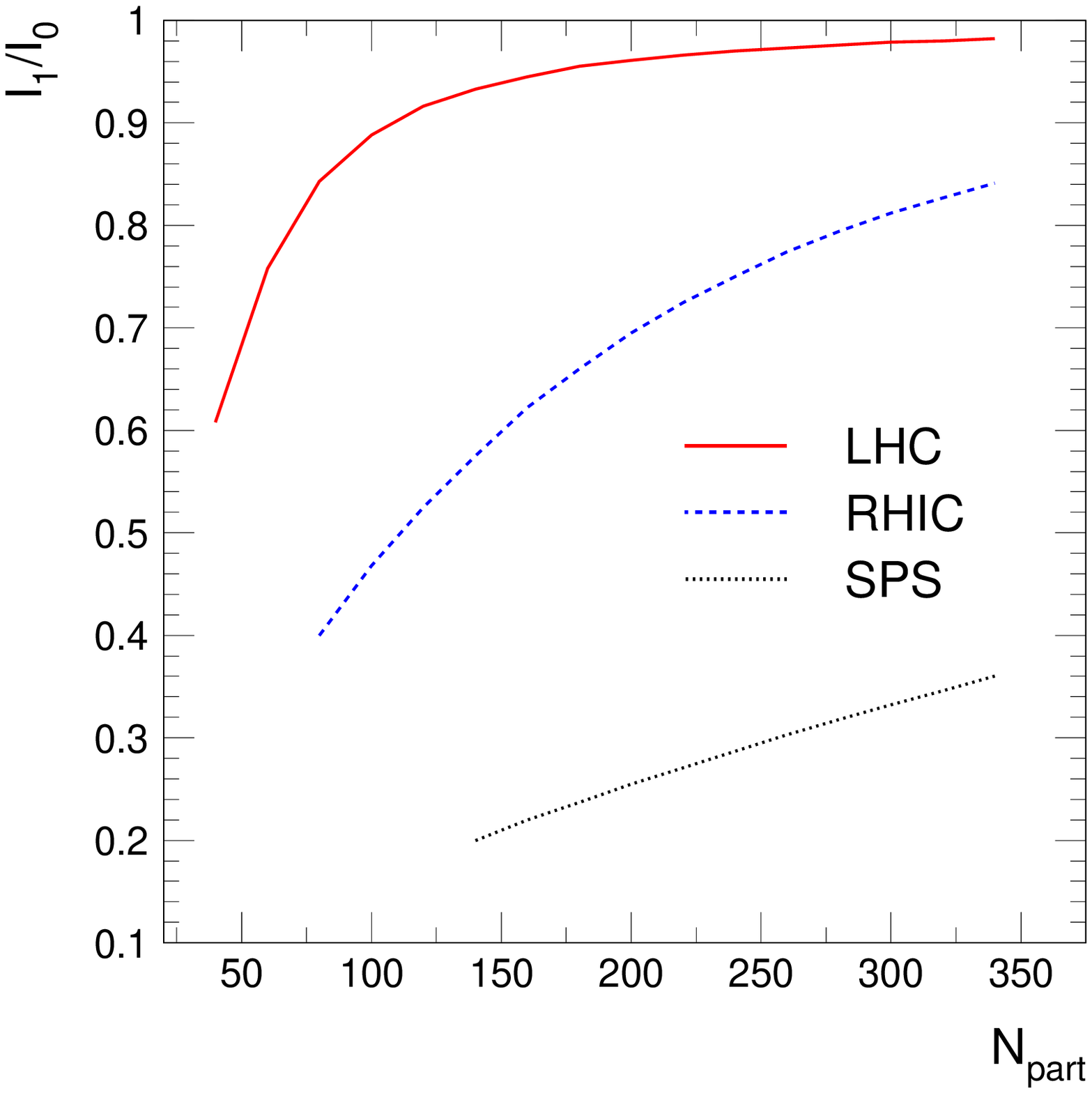}
\end{minipage}  & \begin{minipage}{.49\textwidth}
\vspace{-1cm}
\hspace{-0.5cm}\includegraphics[width=1.15\textwidth]{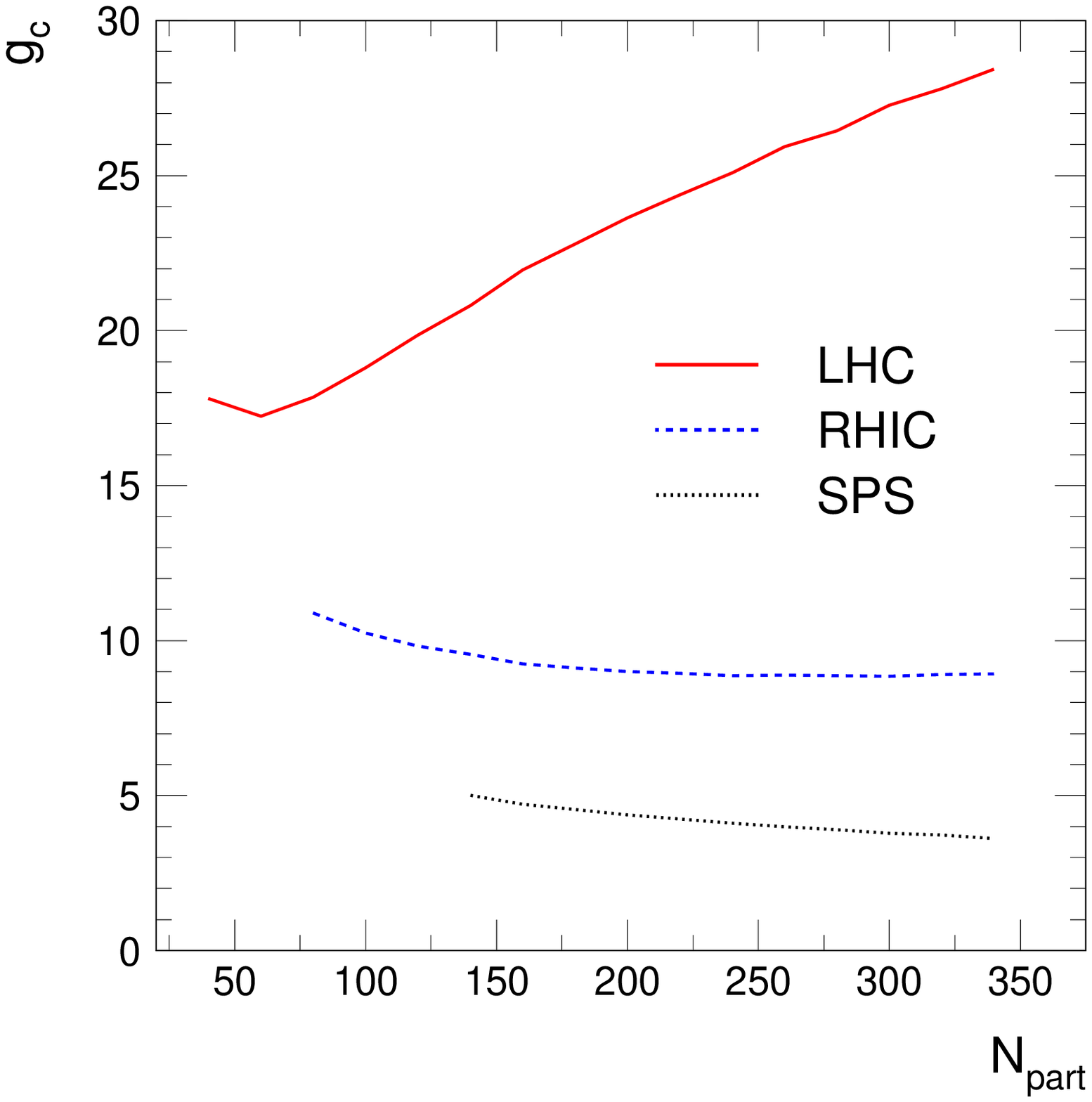}
\end{minipage} \end{tabular}
\caption{Centrality dependence of the canonical suppression for charm, 
$I_1/I_0$ (left panel) and of the charm quark fugacity, $g_c$ (right panel) 
for SPS, RHIC and LHC energies.} 
\label{aa_fig0a}
\end{figure} 

To illustrate some essential features of the model, we show in
Fig.~\ref{aa_fig0a} the centrality dependence of the canonical suppression
factor, $I_1/I_0$, and of the charm quark fugacity, $g_c$, for SPS, RHIC and
LHC energies. The inputs for the calculations are discussed below.  It is
clear that at SPS and RHIC energies, where the number of c $\bar c$ pairs
especially for more peripheral collisions is not much larger than 1, the
centrality dependence of the $J/\psi$ yield, controlled by the $g_c$
parameter, is determined by the canonical suppression. At LHC energy this is a
rather weak effect, owing to a much larger charm production cross section and
to a larger volume at freeze-out (hadronization).  We note that an important
check of the validity of the statistical hadronization model is the
$\psi'/\rm{J}/\psi$ yield ratio.  The model predictions agree with the earlier
data at the SPS energy \cite{pbm1} and we will discuss the present status in
Section \ref{s_cy}.  The model has the following physical input parameters: i)
the temperature, $T$, and baryochemical potential, $\mu_b$, for statistical
model calculations; ii) the heavy quark production cross section in
nucleon-nucleon interactions; iii) the volume of one unit of rapidity,
$V_{\Delta y=1}$ at chemical freeze-out.  We discuss them below, with special
emphasis on the updates used here relative to our previous set of predictions
\cite{aa1}.

Based on our recent analysis of hadron yields within the thermal model 
\cite{aa2}, we use here ($T$,$\mu_b$) of (161,22.4) MeV for RHIC
and (161,0.8) MeV for LHC. 
The uncertainty in $T$ is 4 MeV and we have shown previously \cite{aa1} 
that  the outcome of SHM results exhibit little sensitivity to the precise
value of $T$.
Note that the thermal parameters have been determined from fits of
data at midrapidity in central collisions. From the measured particle 
abundancies \cite{st1}, the centrality dependence of $T$ and $\mu_b$ 
at RHIC is expected to be weak. 
Based on the measured rapidity dependence of particle ratios at RHIC 
\cite{brahms} it is expected that ($T$,$\mu_b$) will be different away 
from midrapidity (in particular, $\mu_b$ is expected to increase), but 
this has a rather small influence on the model results in the present context.

In order to have accurate calculations, it is important to include the
complete spectrum of open charm states.  Based on the latest updates of the
Particle Data Group \cite{pdg}, compared to our previous results \cite{aa1},
we have added for the present calculations 10 new charmed mesons
($D^*_0(2400)$, $D_1(2420)^\pm$, $D^*_{s0}(2317)$, $D_{s1}(2460)$) and 12 new
charmed baryons ($\Sigma_c(2520)$, $\Xi_c^0(2646)$, $\Xi_c'^{+,0}$) plus their
anti-particles.  We now have in the code (including antiparticles) 40 charmed
meson states and 32 charmed baryon states.  As a consequence of these updates,
the J/$\psi$ yield is reduced by 12\% at RHIC and by 16\% at LHC energy as
compared to the results obtained with the previous, less complete, spectrum.
Even if the spectrum of charmed hadrons may still not be complete at this
stage, the effect of further missing states on the J/$\psi$ yield is
suppressed because of their mass. 

To get a feeling for the size of the effect we have investigated the following
scenario: some of the above new meson resonances are
interpreted \cite{maciek} as chiral partners of the known charmed hadrons.
Within this approach, their mass is about 300 to 400 MeV larger than for the
"ordinary" charmed hadrons. It is easy to estimate that, in a thermal
approach, the abundancy of all possible still missing charm states is about
10\% of the abundancy of "ordinary" charmed hadrons, which would be equivalent
to a reduction of the charm production cross section in proportion.  As we
discuss below, the uncertainty in the charm production cross section is much
larger and dominates the uncertainty of our model predictions.  

The spectrum of open bottom mesons and baryons is poorly known in comparison
to the charm sector.  To partially overcome this difficulty, we have added, 
besides the known states \cite{pdg}, excited states with masses derived via 
an analogy with the charm mesons and baryons. Despite this, we should clearly
state that our predictions on $\Upsilon$ production, which we pursue here 
for the LHC energy, are an upper bound.

\begin{figure}[htb]
\begin{tabular}{lr}
\begin{minipage}{.49\textwidth}
\vspace{-1cm}
\centering\includegraphics[width=1.15\textwidth]{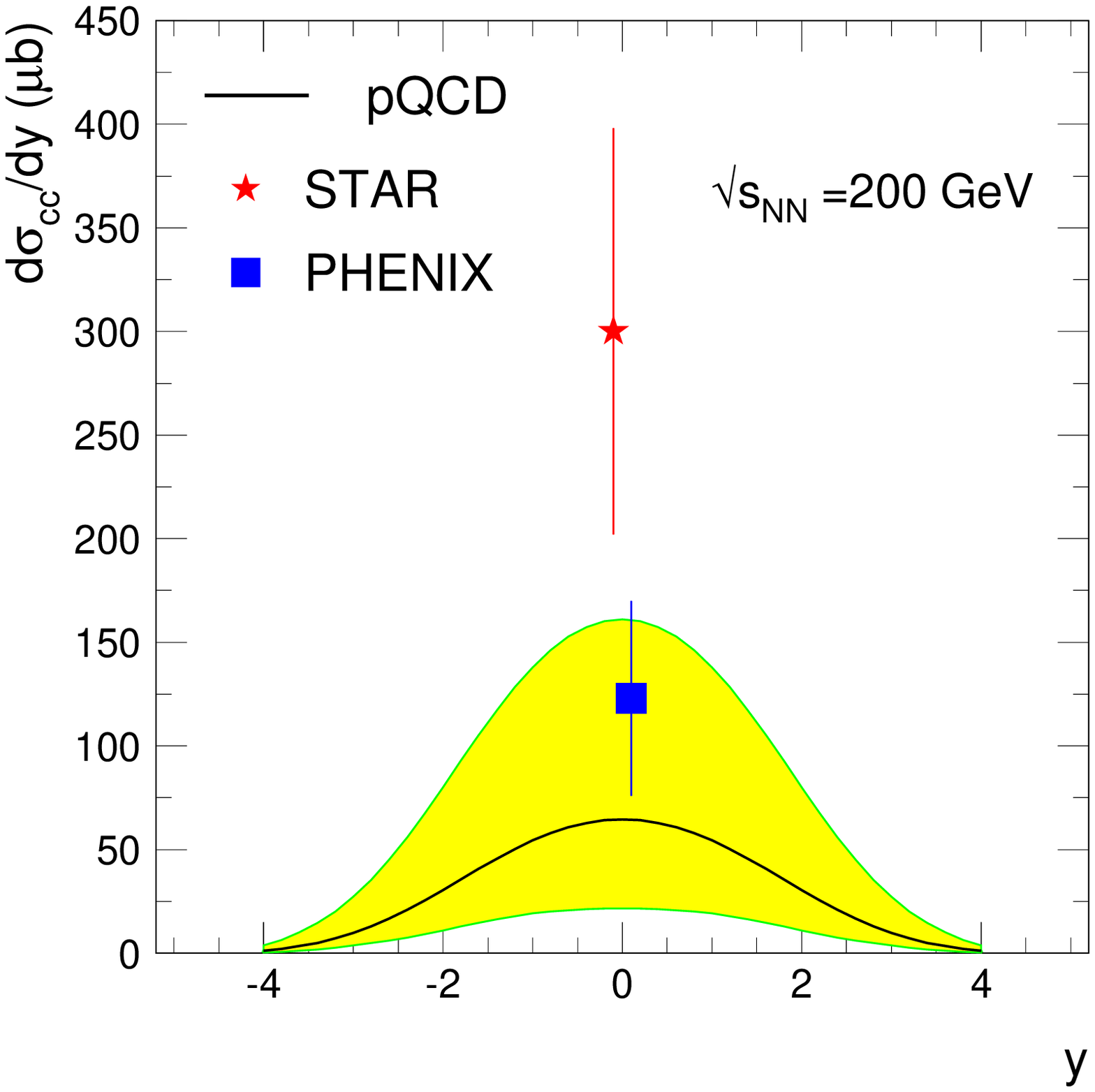}
\end{minipage}  & \begin{minipage}{.49\textwidth}
\vspace{-1cm}
\centering\includegraphics[width=1.15\textwidth]{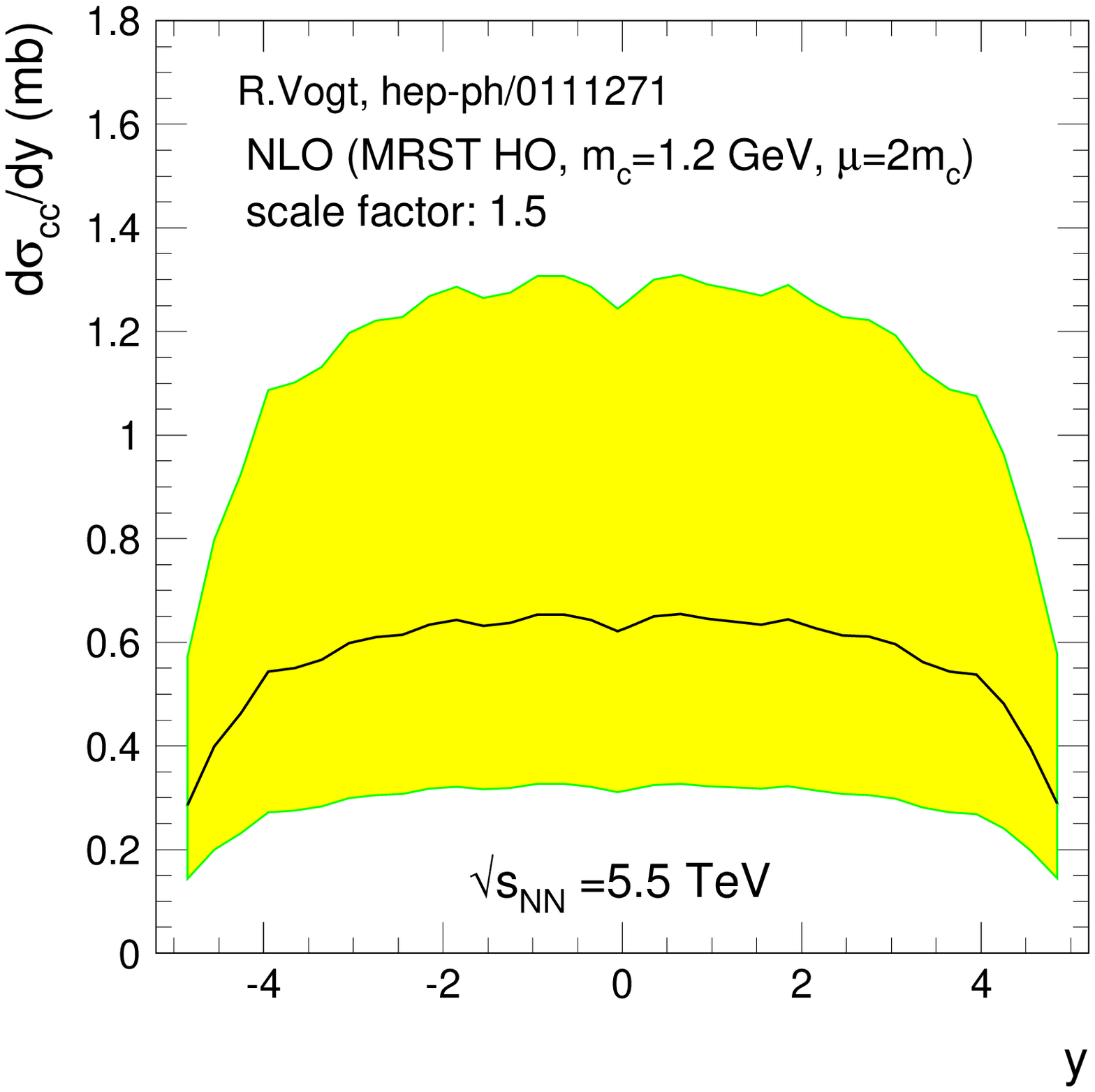}
\end{minipage} \end{tabular}
\caption{Rapidity dependence of the charm production cross section in 
pp collisions. 
Left panel: for RHIC energy, pQCD calculations \cite{cac} and experimental 
values \cite{cc1,cc2}. Right panel: for LHC energy, NLO calculations of 
ref. \cite{rv1} scaled up by a factor of 1.5.}
\label{aa_fig0}
\end{figure} 

The charm production cross section in nucleon-nucleon interactions is, in
absence of direct measurements, taken from perturbative QCD (pQCD)
calculations.  For SPS energy we use the rapidity density of the charm cross
section of ref. \cite{rv1}.  For the RHIC energy, we have adopted the recent
pQCD calculations \cite{cac}, which include a careful estimate of the
theoretical systematic errors.  It was shown earlier \cite{rv1} that, at RHIC
energy, shadowing of the parton distribution functions (PDF) in AA collisions
is a minor effect.  For LHC energy we have adopted the NLO calculations of
ref.~\cite{rv1}, where shadowing for AA collisions is included.  In these
calculations the intrinsic $k_T$ broadening of partons ($\langle
k_T^2\rangle$=1 GeV$^2$) is supplemented by a nuclear broadening of 0.7
GeV$^2$ ($A$=200).  Further, we have scaled up these pQCD calculations by a
factor of 1.5 to bring them in agreement with more recent results \cite{cac1}
(see discussion in \cite{ppr1}).  Following the study made in \cite{cac} we
have assumed the systematic errors of the cross section to be a factor 2 for
SPS and LHC energy. From the cross section, we calculate the number of
produced $c\bar{c}$ pairs as: $N_{c\bar{c}}=\sigma_{c\bar{c}}\cdot T_{AA}$,
where $T_{AA}$ is the nuclear overlap function, calculated using \cite{dar}.

The rapidity distributions of the charm production cross section in pp
collisions for RHIC and LHC energies are shown in Fig.~\ref{aa_fig0}.  As seen
in the left panel in Fig.~\ref{aa_fig0}, at RHIC the extracted experimental
charm production cross section at midrapidity of PHENIX \cite{cc2} of $123\pm
47~\mu$b is, within the errors, in agreement with the pQCD calculations of
$\ud \sigma_{c\bar{c}}^{pQCD}/\ud y=63.7^{+95.6}_{-42.3}~\mu$b \cite{cac}.
The measured value of $300\pm 98~\mu$b by STAR \cite{cc1} is significantly
larger.  Note that the experimental values are extracted mostly from single
electron spectra \cite{cc1,cc2}, a difficult measurement which is reflected in
the size of the errors. It was shown \cite{cac} that the PHENIX single
electron transverse momentum ($p_t$) spectra \cite{cc2b} are compatible with
the upper limit of the systematic errors for the pQCD calculations, as also
seen in Fig.~\ref{aa_fig0} for the $p_t$-integrated value.  We note that, at
Tevatron, the measured cross section of D mesons \cite{cdf} was underpredicted
by pQCD calculations \cite{cac2} by up to a factor of 2 for the lowest
measured transverse momentum ($\simeq 5$ GeV/c).  However, in more recent
calculations \cite{kni} this discrepancy is reduced to a factor of 1.5, so
that now data and theory are compatible within errors.

\begin{figure}[htb]
\begin{tabular}{lr}
\begin{minipage}{.49\textwidth}
\vspace{-1cm}
\centering\includegraphics[width=1.15\textwidth]{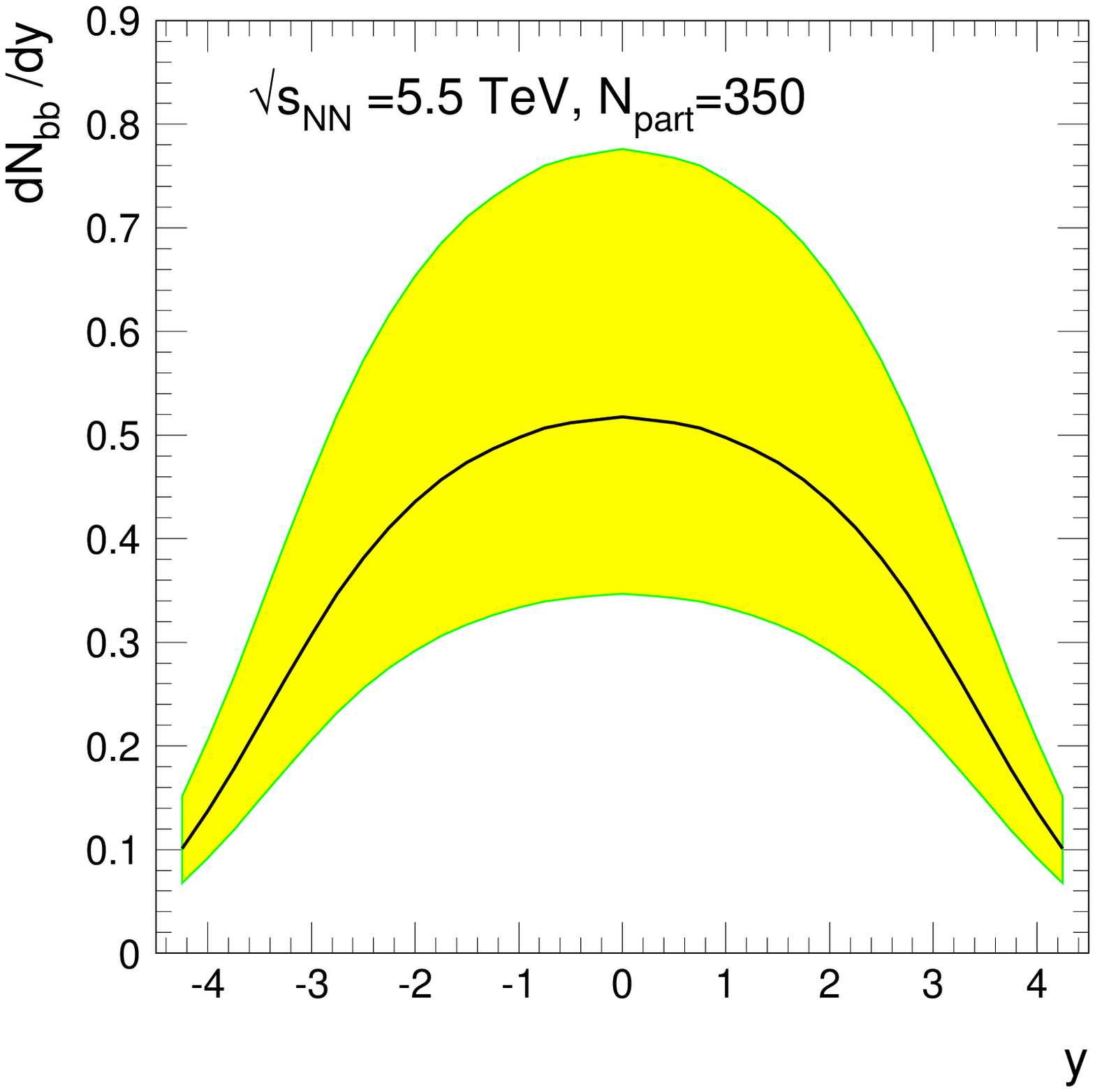}
\end{minipage}  & \begin{minipage}{.49\textwidth}
\vspace{-1cm}
\centering\includegraphics[width=1.15\textwidth]{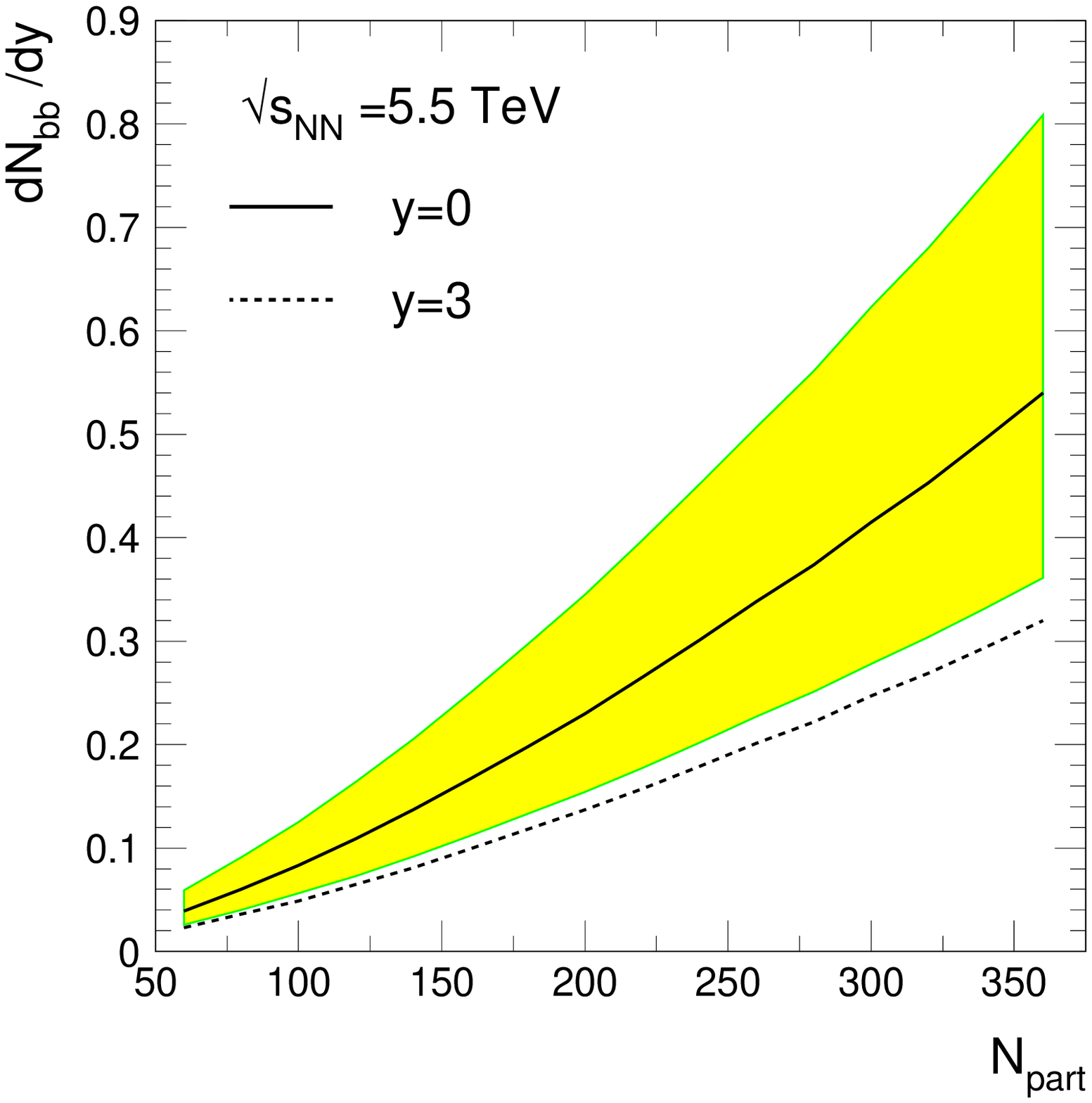}
\end{minipage} \end{tabular}
\caption{Rapidity (left panel, for central collisions) and centrality 
dependence (right panel) of the number of $b\bar{b}$ pairs per unit 
of rapidity. The lines are for the central value of the bottom
cross section \cite{rv1}, the shaded areas are bounds for the uncertainty.}
\label{aa_fig8}
\end{figure} 

For bottom production cross sections we use the pQCD calculations of ref.
\cite{rv1}.
For the uncertainty of the cross section we assume a factor of 1.5 up and down.
In Fig.~\ref{aa_fig8} we show the bottom quark pair production rates
as a function of rapidity for central Pb-Pb collisions and as a function of
centrality for $y$=0 and $y$=3.

In our approach all charmonia are formed by $c\bar{c}$ recombination at the
phase transition. An interesting question is the volume within which
recombination can take place. A natural size for this quantity is the volume
corresponding to a slice of one unit of rapidity.  We have shown earlier
\cite{aa1} that, at RHIC, the dependence of the results on the magnitude of
the rapidity width is small, as long as intervals of 1-3 units of rapidity
are considered.  This dependence is even smaller for the LHC case.  The volume
$V_{\Delta y=1}$ is obtained from the calculated thermal densities and the
experimental values of charged particles rapidity densities, $\ud N_{ch}/\ud
y$, at midrapidity.  For central collisions (number of participating nucleons
$N_{part}$=350) we obtain at SPS $V_{\Delta y=1}$=1200 fm$^3$ , while at RHIC,
$\ud N_{ch}/\ud y$=701 leads to $V_{\Delta y=1}$=2400 fm$^3$ \cite{aa2}.  At
LHC, based on a recently-proposed parametrization \cite{asw} of the energy
dependence of $\ud N_{ch}/\ud y$, the expected value is $\ud N_{ch}/\ud
y$=1816, leading to $V_{\Delta y=1}$=6200 fm$^3$ at midrapidity.  This
presently used value of $\ud N_{ch}/\ud y$ at LHC is slightly lower than our
previous phenomenological extrapolation \cite{aa1}.  We have shown earlier
\cite{aa1} that the uncertainty in the volume is affecting the model
predictions rather marginally.  Our previous results \cite{aa1} were obtained
assuming a linear scaling of $\ud N_{ch}/\ud y$ (and consequently 
$V_{\Delta y=1}$) with $N_{part}$.  
For the present calculations the scaling is done for the 'core' region only 
(see below).

\begin{figure}[ht]
\centering\includegraphics[width=.65\textwidth]{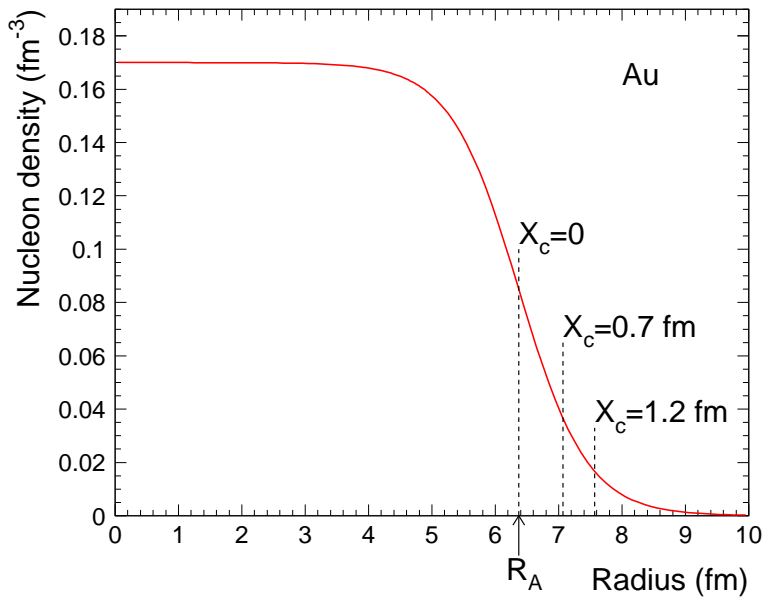}
\caption{The nuclear density (Woods-Saxon) distribution for the Au nucleus.
The core corresponds to the radius $R_A+X_c$, beyond this radius the 
nucleons are in the corona region.} 
\label{aa_fig1b}
\end{figure}

\begin{figure}[ht]
\hspace{-.7cm}
\begin{tabular}{lr}
\begin{minipage}{.49\textwidth}
\vspace{-1cm}
\centering\includegraphics[width=1.2\textwidth]{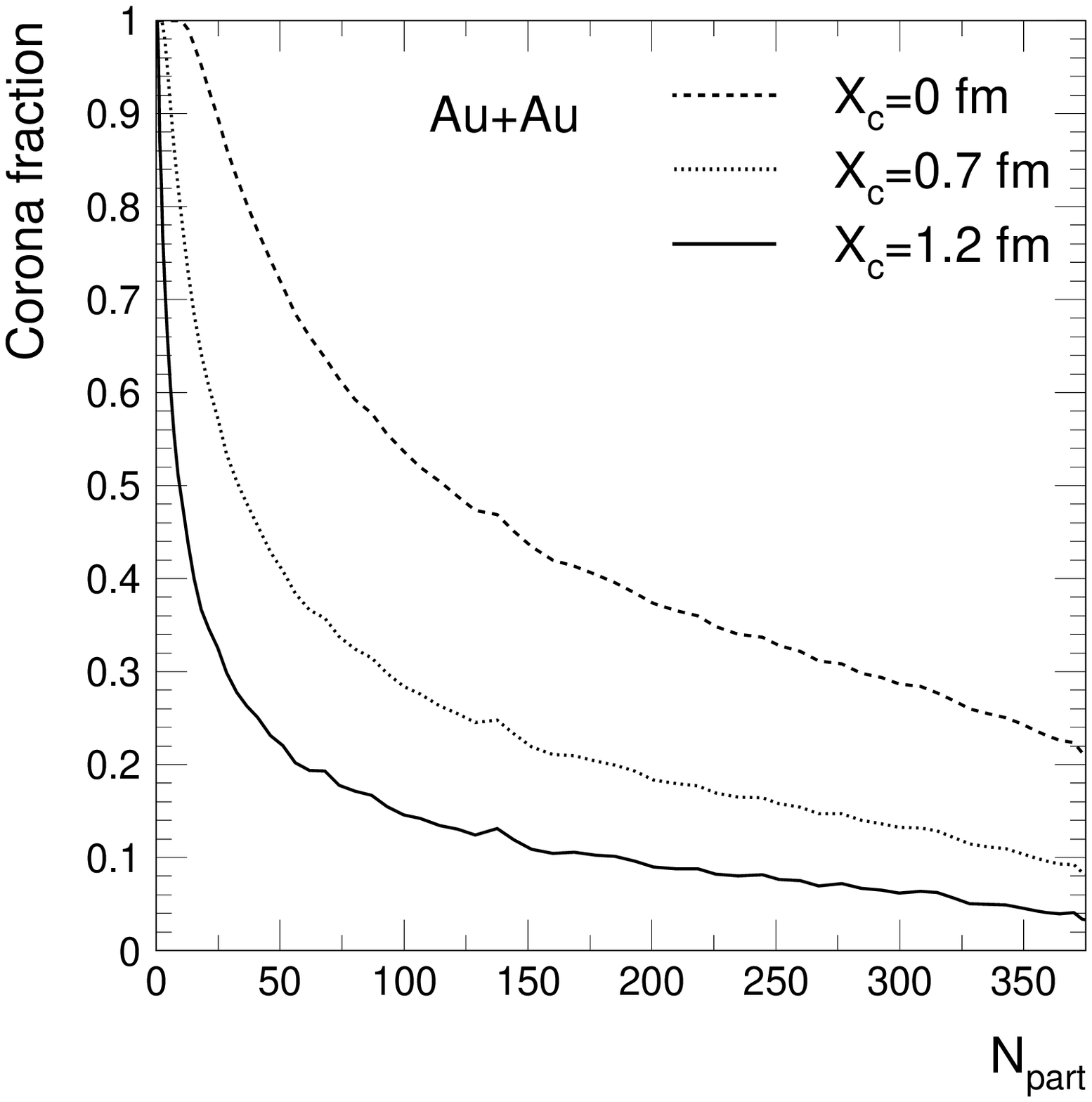}
\end{minipage}  & \begin{minipage}{.49\textwidth}
\vspace{-1cm}
\centering\includegraphics[width=1.2\textwidth]{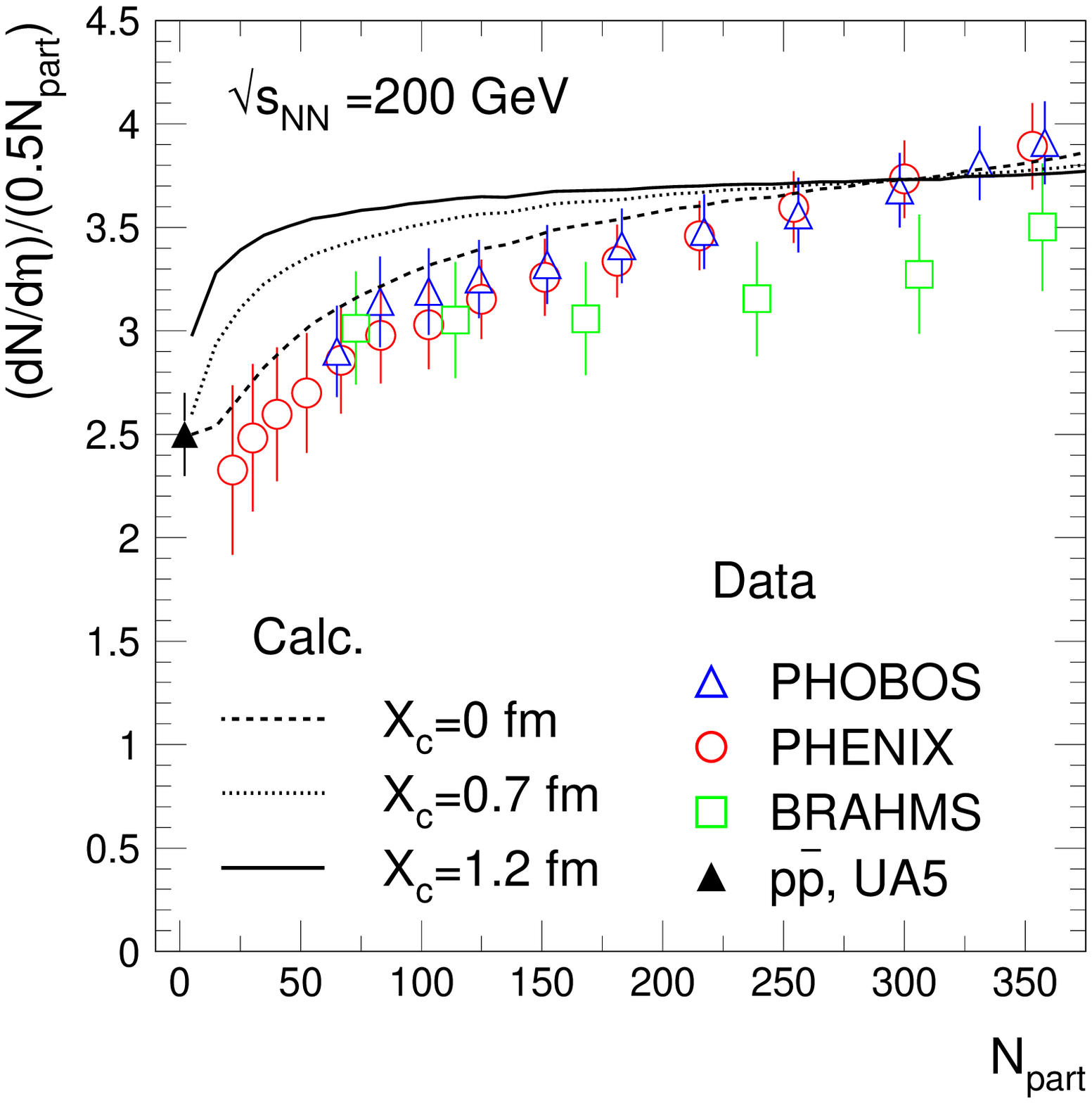}
\end{minipage}
\end{tabular}
\caption{Left panel: the fraction of nucleons in corona as a function 
of centrality. 
Right panel: the centrality dependence of charged particle yields at 
midrapidity. The data \cite{phe2} are compared to estimates based on 
the corona contribution (the curves are arbitrarily normalized).}
\label{aa_fig1}
\end{figure}

{\it 
An important effect to be included when studying the centrality dependence in
nucleus-nucleus collisions is the so-called "corona effect". Nucleons from the
surface of the colliding nuclei do not participate in the formation of the hot
fireball (core) were QGP is assumed to be produced. This fact has recently
been emphasized also in ref.~\cite{wer}. Since the SHM applies only to the QGP
zone, and since charmonium production in nucleon-nucleon collisions is, in
general, very different from that predicted in the SHM, it is relevant to
distinguish between core and corona in our approach.  To quantify core and
corona, we use calculations of the nuclear overlap \cite{dar}, employing a
Woods-Saxon nuclear density distribution, plotted for the Au nucleus in
Fig.~\ref{aa_fig1b}.  We define the core region as corresponding to the
half-density nuclear (charge) radius ($R_A$=6.37 fm for the Au nucleus)
extended by a thickness $X_c$ (see Fig.~\ref{aa_fig1b}) and derive in this
manner $N_{part}$ and $N_{coll}$ for the core and corona regions.  The
fraction of participating nucleons contained in the corona region is shown as
a function of centrality in the left panel of Fig.~\ref{aa_fig1} for three
values of $X_c$. Effectively this three-dimensional approach contains a 
sharp transition between core and corona. We treat the core as QGp using the
SHM and the corona as superposition of nucleon-nucleon collisions.
}

It was recently demonstrated \cite{wer} that the centrality dependence of 
particle production at RHIC and SPS can be well explained assuming an 
interplay of bulk production in the core and of elementary collisions 
in the corona part.
Within this picture, we compare in the right panel of Fig.~\ref{aa_fig1} our
expectation for the centrality dependence of charged particle multiplicity
with RHIC data from PHOBOS \cite{pho}, PHENIX \cite{phe2} and BRAHMS
\cite{brah}.  For the elementary collisions we use $\ud N_{ch}/\ud \eta$=2.5
(at midrapidity, non single-diffractive), measured in p$\bar{\mathrm{p}}$
collisions at $\sqrt{s}$=200 GeV \cite{ua5}.  
We normalize our calculations to the measured value \cite{phe2} for 
$N_{part}$=300 to extract for the core $\ud N_{ch}/\ud \eta/(N_{part}/2)$=
4.23, 3.92 and 3.81 for $X_c$= 0, 0.7 and 1.2 fm, respectively.  
We conclude that this treatment of the corona effect reproduces the 
trend in the data only for the case $X_c$=0 which we consider somewhat 
extreme since, effectively, pA collisions do not contribute to core particle
production (see above). We note also that, using BRAHMS data, one may come to
a somewhat different 
conclusion. 
We consider for the corona thickness $X_c$=1.2 fm as a realistic
value, since at that point the density has dropped to about 10 \% of the
central value and  adopt it  as a baseline choice for our following model 
calculations.  
In this case our calculations account for the measured
centrality dependence of bulk hadron production only partially.
On the other hand, the interpretation of such data within saturation 
("color glass condensate") models \cite{kha,asw} is quite successful.  
However, it is clear from the present exercise that the corona
contribution needs to be considered  in order to have a realistic modelling
of the collision \cite{wer}. To let the reader judge the influence of the
corona effect we provide calculations for the centrality dependence of
J/$\psi$ production with and without considering the corona.  
 
\begin{figure}[ht]
\begin{tabular}{lr}\begin{minipage}{.48\textwidth}
\vspace{-1cm}
\hspace{-0.5cm}\includegraphics[width=1.2\textwidth]{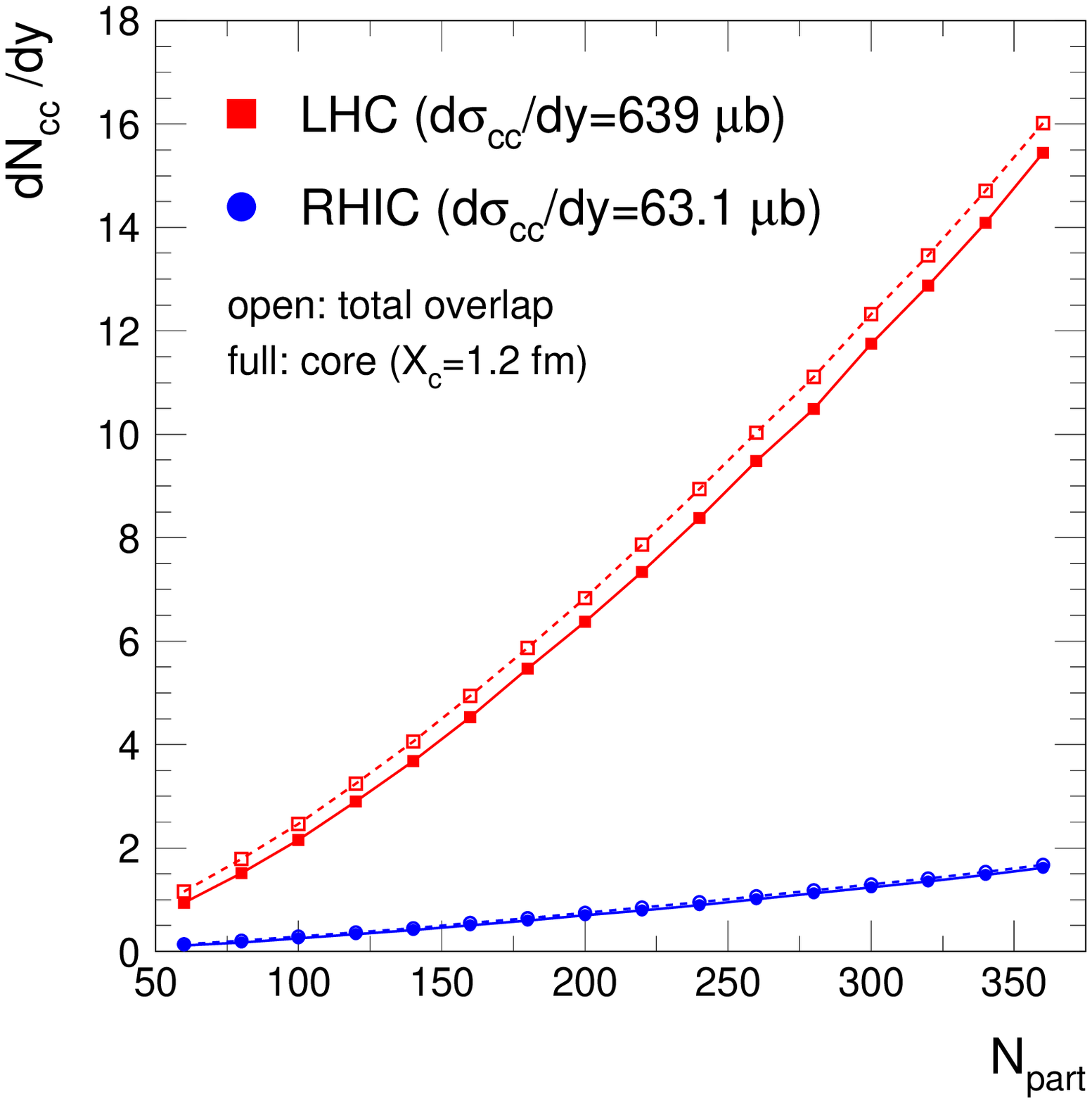}
\caption{Centrality dependence of rapidity density of charm yields at
RHIC and LHC energies for the total overlap (open symbols) and for core only
(full symbols).} \label{aa_fig2a}
\end{minipage}  
& 
\begin{minipage}{.48\textwidth}
\vspace{-1cm}
\hspace{-0.5cm}\includegraphics[width=1.2\textwidth]{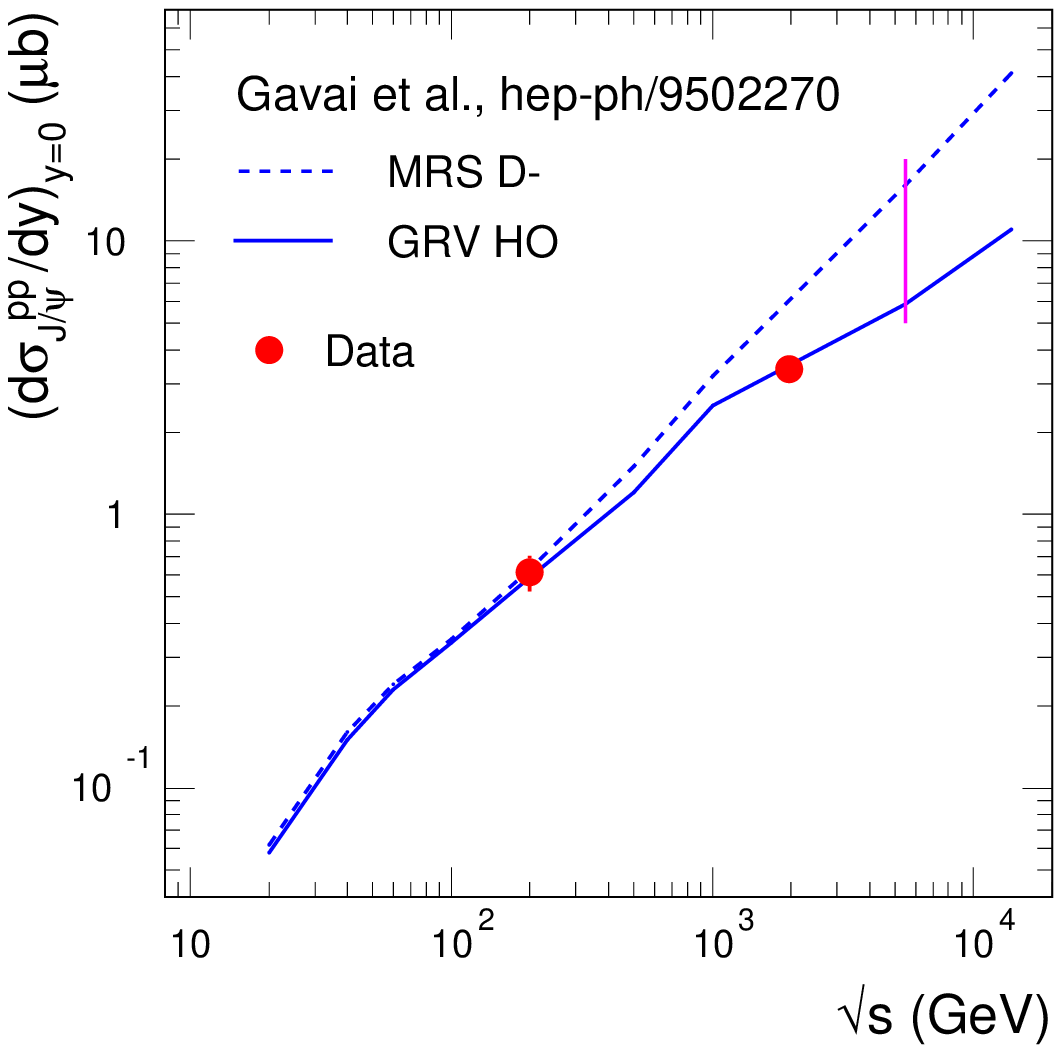}
\caption{Energy dependence of rapidity density of $J/\psi$ cross section 
in pp($\bar{\mathrm{p}}$) collisions (see text for details).
The LHC energy for Pb+Pb is marked by the vertical line.} \label{aa_fig2b}
\end{minipage} \end{tabular}
\end{figure}

Based on the central values of the pQCD charm production cross sections 
\cite{cac,rv1}, we show in Fig.~\ref{aa_fig2a} the centrality dependence 
of the number of $c\bar{c}$ pairs for central rapidity for the RHIC 
and LHC energies. In the following we perform calculations within the 
statistical hadronization model for the core, and add the contribution 
of the corona region for which we use the $J/\psi$ cross section in
elementary collisions.  The energy dependence of the $J/\psi$ cross
section ($\ud \sigma/\ud y$) in pp collisions is shown in
Fig.~\ref{aa_fig2b}.  Recent experimental data at RHIC from PHENIX
\cite{phe3} and at the Tevatron from CDF \cite{pau} are compared
to calculations by Gavai et al. \cite{gav} for two PDFs. In our model
calculations we employ for RHIC the measured \cite{phe3} $J/\psi$
cross section of 0.724 $\mu$b, which is about 1\% of the pQCD charm
production cross section \cite{cac}.  Assuming the same 1\% of the
calculated pQCD charm cross, for LHC we derive a $J/\psi$ cross
section of 6.39 $\mu$b, which we use in the following calculations.
This cross section is in line with the calculations \cite{gav} which
reproduce the Tevatron data \cite{pau} (see Fig.~\ref{aa_fig2b}).
For  SPS energy we employ a $J/\psi$ cross section at midrapidity 
in pp collisions of 50 nb, obtained from an interpolation based on 
a recent compilation of data \cite{herab}. These values are summarized 
in Table~\ref{tab1}.

\vspace{0.5cm}
\begin{table}[h]
\caption{Summary of the values used in our calculations for charm.
The values for $\ud \sigma_{c\bar{c}}^{pp}/\ud y$ are from pQCD calculations,
the values for $\ud \sigma_{J/\psi}^{pp}$ are from measurements, either
interpolated (for the SPS energy, cf. ref. \cite{herab}), or directly measured 
at RHIC \cite{phe3} or assumed as 1\% of $\ud \sigma_{c\bar{c}}^{pp}/\ud y$ 
at LHC energy.} 
\label{tab1}
\begin{tabular}{l|cc}
$\sqrt{s_{NN}}$ (GeV) & 
$\ud \sigma_{c\bar{c}}^{pp}/\ud y$ ($\mu$b) &
$\ud \sigma_{J/\psi}^{pp}/\ud y$ ($\mu$b) 
\\ \hline
17.3 & 5.7$^{+5.7}_{-2.8}$ & 0.050$\pm$0.030 \\
200  & 63.7$^{+95.6}_{-42.3}$  & 0.774$\pm$0.124 \\
5500 & 639$^{+639}_{-319}$ & 6.4$\pm$3.2 \\
\end{tabular}
\end{table}

The yield of J/$\psi$ mesons, e.g., is then calculated as
follows:
\be
N_{J/\psi} = N_{J/\psi}^{core} + N_{J/\psi}^{corona}
\ee
where 
\be
N_{J/\psi}^{core} = g_c^2 n_{J/\psi}^{th} V^{core}
\ee
and
\be 
 N_{J/\psi}^{corona} = N_{coll}^{corona}
 \sigma_{J/\psi}^{pp}/\sigma_{inel}^{pp}.
\ee
and we use for $\sigma_{inel}^{pp}$ values of 30, 42, and 60 mb at SPS, RHIC,
and LHC energies, respectively.

The $\Upsilon(1S)$ cross section in pp collisions at the LHC energy
is assumed to be 36.8 nb, which is 1.7$\cdot 10^{-3}$ of the $b\bar{b}$ 
cross section at $\sqrt{s_{NN}}$=5.5 TeV. 
This is the fraction derived from the Tevatron data at 1.8 TeV 
\cite{pau,cac3,cac4}.

\begin{figure}[ht]
\vspace{-1cm}
\centering\includegraphics[width=.7\textwidth]{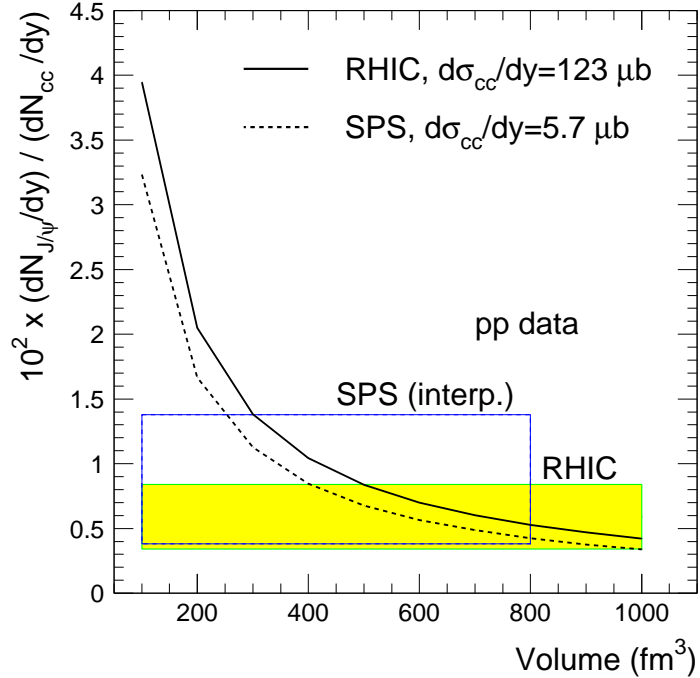}
\caption{Volume dependence of the rapidity density of the  $J/\psi$ yield
  calculated within the SHM,
normalized to the $c\bar{c}$ yield. The measured value for the RHIC energy 
\cite{phe3,cc2} is indicated by the shaded band, while the open box 
depicts the interpolated value for the SPS energy.} \label{aa_figx}
\end{figure}

Before proceeding to discuss our calculations for nucleus-nucleus 
collisions, we investigate the model predictions for the proton-proton 
case. In Fig.~\ref{aa_figx} we show the volume dependence of the predicted 
$J/\psi$ yield normalized to the $c\bar{c}$ yield.
It is seen that, for a realistic volume expected for pp collisions, 
of about 30-50 fm$^3$, the model strongly overestimates the measured 
\cite{phe3,cc2} or interpolated \cite{herab} values for the RHIC and SPS 
energies, respectively.
This is expected, as charm quarks are not likely to thermalize for small 
volumes.
As a consequence, we consider, somewhat schematically, the volume for which
the model calculations reach the measured values in pp collisions to be 
the minimal core volume required for the validity of our model. 
Based on the results shown in Fig.~\ref{aa_figx}, we adopt the value 
$V_{QGP}^{min}$=400 fm$^3$. 
Below this value, we consider that the $J/\psi$ production takes place 
exclusively via hard processes in elementary collisions.
In nature, the transition from the QGP region to the pp region is likely 
to be continuous, but we have no quantitative  means to model this transition.
Also, from Fig.~\ref{aa_figx} it is clear that $V_{QGP}^{min}$ carries 
a rather large uncertainty.  
In addition, we do not consider the experimental fluctuations in the 
centrality selection which would further smoothen the transition.

\section{Annihilation of charm quarks in the QGP} \label{s_anni}

In this section we investigate the assumption, central to our model, that the
number of $\bar{\rm{c}}$ and c quarks which are produced in initial hard
collisions stays constant during the evolution of the QGP until the
critical temperature T$_c$ is reached and the system hadronizes. For such an
estimate we assume that the light quarks and gluons reach equilibrium to form
a QGP relatively quickly after the first hard collisions, on a time scale of
the order of 1 or at most a few fm/c. We first note that further production of
$\bar{c}c$ pairs in the QGP can be completely neglected at temperatures
$T< 1$ GeV \cite{redlich_pbm}, as the charm quarks are far from chemical
equilibrium. On the other hand, charm quark annihilation may take place via
reactions of the type 
\be
c + \bar{c} \rightarrow g + g
\label{e3-1}
\ee
 or
\be
c + \bar{c} \rightarrow q + \bar{q},
\label{e3-2}
\ee or via more complex reactions such as those forming 3 or more gluons, a
quark-antiquark pair plus gluons, etc. Within the framework of perturbative
QCD, the cross sections for Eqs.~\ref{e3-1},~\ref{e3-2} can be evaluated.  In
the following, we perform such an evaluation for Eq.~\ref{e3-1} and use it to
determine the annihilation rate of charm quark pairs in the QGP via this
channel.  The 2-gluon annihilation is the dominant channel, its cross
section significantly exceeding all other listed cross sections
\cite{nachtmann}.

Channels with 3 or more gluons are not expected to play a major role in our
estimate. An estimate of the relative size of annihilation into 3 gluons vs. 
2 gluons can be obtained by inspecting the width of the two lowest-lying 
charmonia $\eta_c$ and J/$\psi$. Because of  C-parity conservation the 
$\eta_c$'s width is dominated by 2-gluon decay while J/$\psi$ decays mainly 
into 3 gluons. In fact, their measured widths are 25 and less than 0.1 MeV,
respectively.  To leading order in $\alpha_s$ and in non-relativistic
approximation, the ratio of these widths 
is:
\be
\frac{\Gamma(J/\psi)}{\Gamma(\eta_c)} = \frac{4}{9\pi} (\pi^2-9) \alpha_s.
\ee
Evaluation of this expression for $\alpha_s(m(J/\psi)^2) \approx 0.3$
\cite{bethke} 
overestimates the measured ratio by approximately an order of
magnitude. Annihilation into 3 gluons is obviously strongly suppressed 
relative to the 2-gluon process. For more information on this issue see
\cite{eichten}. Moreover, multi-gluon annihilation should be further 
suppressed by phase space if gluons acquire, as expected, a thermal mass 
in the QGP. We conclude that annihilation into 2 gluons is the dominant
channel of the inelastic $c\bar c$ cross section. 

To evaluate the total annihilation rate we start from the rate equation
\be
\frac{dr_{c\bar{c}}} {d\tau} = n_c n_{\bar{c}} \langle
\sigma_{c\bar{c}\rightarrow gg} v_r \rangle,
\label{e3-3}
\ee 
where $\langle \sigma_{c\bar{c}\rightarrow gg} v_r \rangle$ is the thermal
average of the annihilation cross section times the relative velocity $v_r$ in
the QGP, and $n_c$ = $n_{\bar{c}}$ is the charm quark density. The quantity
$\frac{dr_{c\bar{c}}} {d\tau}$ is the annihilation rate per volume or the rate
of change of the charm quark density. 
The cross section for Eq.~\ref{e3-1} is taken to first order in the strong
coupling constant $\alpha_s$ from the work of \cite{glu}, where the inverse
cross section for $g + g \rightarrow c + \bar{c}$ is computed, applying
detailed balance. To get an estimate of the upper limit for this cross section
in the QGP we evaluate the equation of \cite{glu} using $\alpha_s = 1$ and for
a charm quark mass of $m_c$ = 1.5 GeV. 

The charm quark density $n_c$ depends on the evolution time
of the QGP, i.e. 
\be
n_c = \frac{\ud N_c/\ud y(\tau)} {V(\Delta y = 1,\tau)} \leq 
\frac{\ud N_c/\ud y(\tau_0)} {V(\Delta y = 1,\tau)},
\label{e3-4}
\ee
where $\tau_0$ is the initial time of QGP formation and $\ud N_c/\ud y$
is the charm quark rapidity density. 

The total annihilation yield of charm quarks in the QGP is then given by
\be
N_{c \bar{c}}^{anni} = \int_{\tau_0}^{\tau_c} \frac{\ud r_{c\bar{c}}}{\ud \tau}
V(\Delta y =1,\tau) \ud \tau .
\label{e3-5}
\ee

To proceed further we need to model the evolution of the QGP.
In the spirit of our upper limit estimate we assume here a 1-dimensional
Bjorken-type expansion of the QGP yielding, 
using entropy conservation, a relation between $T$ and $\tau$: 
\be
\frac{\pi^2}{45} (32+21 N_f)T^3 \tau =3.8\frac{\ud N/\ud y}{A_{\perp}},
\label{e3-6}
\ee
where $N_f$ is the (effective) number of massless flavors, and $\ud N/\ud y$
is the total particle rapidity density.
The transverse system size $A_{\perp}$ is about 150 fm$^2$ at T$_c$ for 
central Au-Au or Pb-Pb collisions.  Using this scenario we get for the 
temporal evolution of the volume then 
\be
V(\Delta y =1,\tau) = A_{\perp} \tau.
\label{e3-7}
 \ee

The total annihilation yield of charm quarks in the QGP is, using 
Eqs.~\ref{e3-4},\ref{e3-5},\ref{e3-7}, given by
\be
N_{c \bar{c}}^{anni}   \leq \left(\frac{\ud N_c}{\ud y}(\tau_0)\right)^2 
\frac{1}{A_\perp} \int_{\tau_0}^{\tau_c} \frac{\ud \tau} {\tau} \langle
\sigma_{c\bar{c}\rightarrow gg} v_r \rangle.
\label{e3-8}
\ee

\begin{figure}[ht]
\begin{tabular}{lr}
\begin{minipage}{.48\textwidth}
\vspace{-1cm}
  \centering\includegraphics[width=1.2\textwidth]{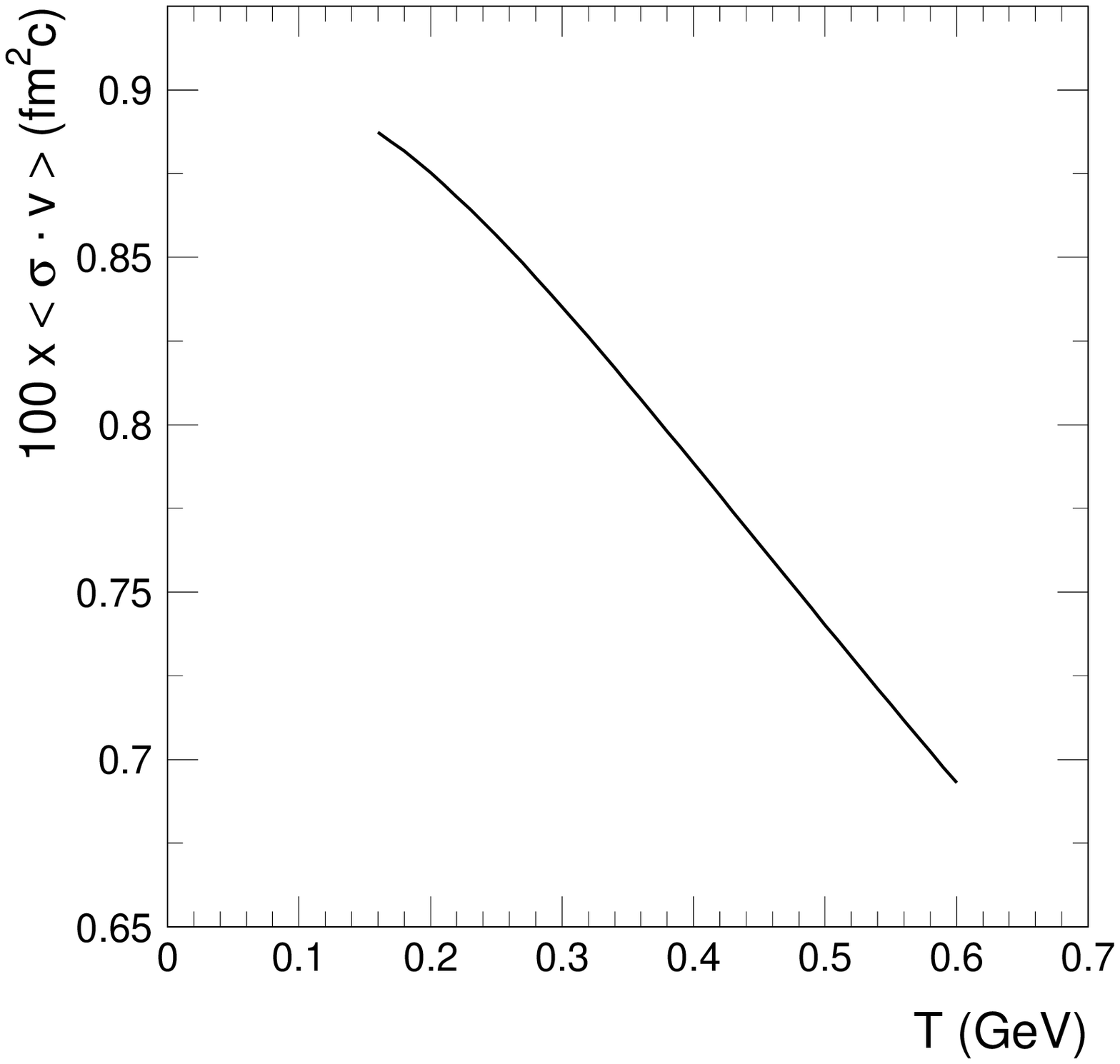}
  \caption{Temperature dependence of the thermal average as defined in
    Eq.~\ref{e3-3} (see text).}
  \label{aa_fx1}
\end{minipage}  & \begin{minipage}{.48\textwidth}
\vspace{-1cm}
  \centering\includegraphics[width=1.2\textwidth]{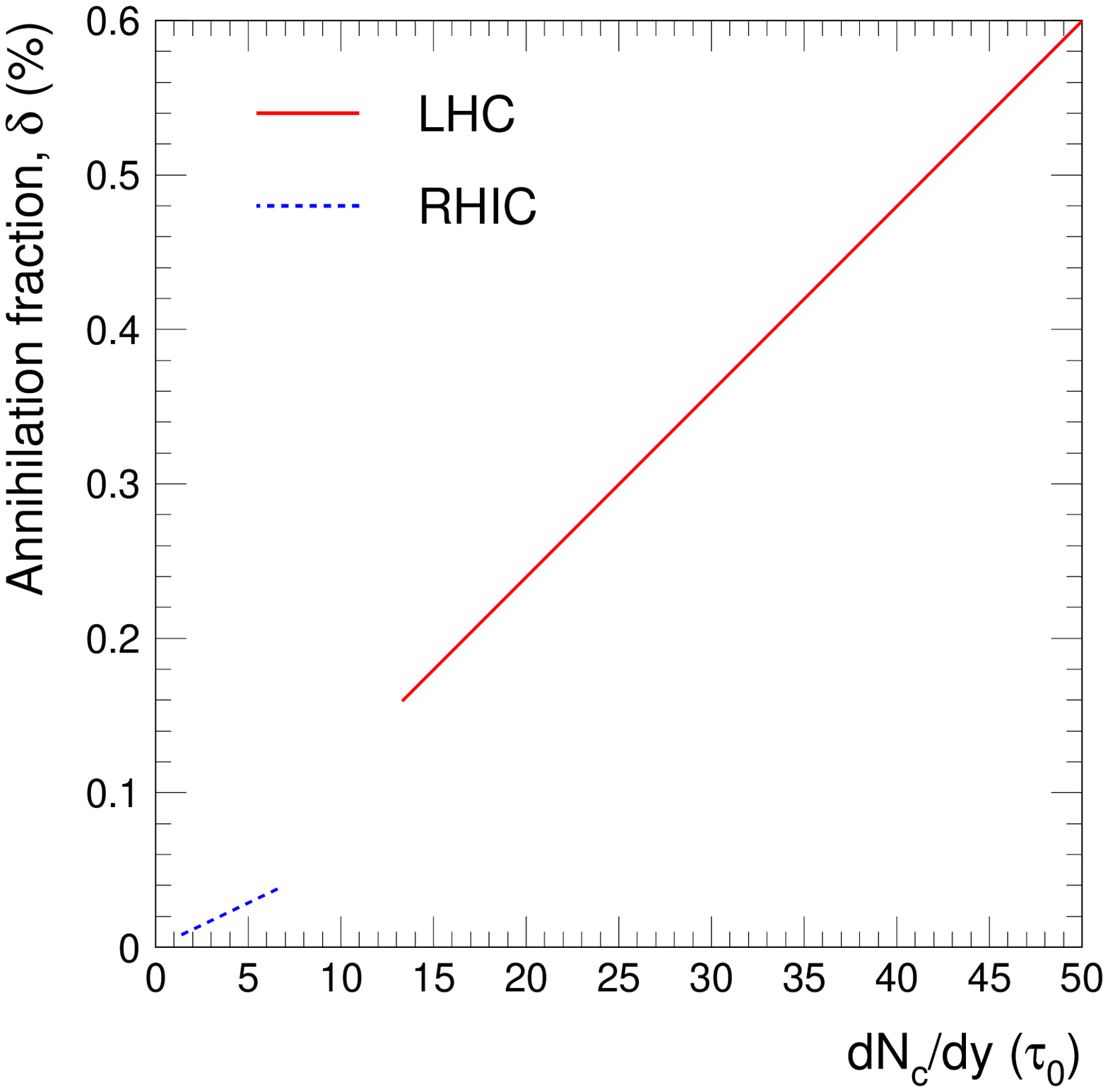}
\caption{Annihilation rate as a function of initial charm rapidity density.}
  \label{aa_fx2}
\end{minipage}
\end{tabular}
\end{figure}

In the following we consider 2 scenarios (assuming N$_f$ = 2.2 and $\tau_0 =
1$ fm for both):
i) at  RHIC energy, this leads to $\tau_c$ = 2.7 fm and to an initial 
temperature $T(\tau_0) = 225 $ MeV  for a charged particle rapidity density 
of 660 and 
ii) the LHC energy scenario results in $\tau_c$ = 8.3 fm and 
$T(\tau_0) = 325 $ MeV for a charged particle rapidity density of 2000.
Following \cite{ko_lin} we evaluate the temperature dependence of the thermal
average in Boltzmann approximation. The resulting temperature dependence of the
thermal average is presented in Fig.~\ref{aa_fx1}. Since the charm quark
annihilation cross section decreases  with increasing scale
($\sqrt{s}$), the resulting thermal average also drops with increasing
temperature. 

To get numerical results on the annihilation rate we determine the integral in
Eq.~\ref{e3-8} for both scenarios 
numerically. Note that the lifetime of the QGP enters (to 1st order) only
logarithmically, implying that the details of the expansion are not very
important for our estimate. In particular, using much smaller values for
$\tau_0$ and correspondingly higher initial temperatures changes the results
only marginally.
We furthermore add that the volume used here for
an upper limit on the annihilation yield is computed when the system reaches
$T_c$. Taking into account the increase of volume until chemical freeze-out,
which is used in the next section,  would further decrease the annihilation
yield and, hence, is not considered here.

The final results of the estimate for the fraction of charm quarks lost via
annihilation, $\delta=N_{c \bar{c}}^{anni}/dN_c/dy(\tau_0)$ are given in 
Fig.~\ref{aa_fx2} for the RHIC and LHC scenarios.
Since the annihilation yields increase quadratically with the
charm quark rapidity density we start, in the spirit of an upper limit
estimate,  the calculation at the values predicted
by pQCD calculations for each scenario, i.e. $\frac{ dN_c}{dy}(\tau_0)
= \frac{dN_{c\bar{c}}^{pQCD}}{dy}$ = 1.4 and 13.3, and then increase these
numbers by up to a factor 5. 

In general, the fraction of annihilating $c \bar{c}$ pairs in the QGP is 
negligible for both scenarios. Even for an unrealistically high value of 
$\frac{ dN_c}{dy}(\tau_0)\approx 60$ in the LHC scenario, the total fraction 
of annihilated pairs is less than 1\%, and decreases linearly for lower 
charm quark densities.
For realistic scenarios our estimate implies that charm quark annihilation in
the plasma can be safely neglected. Along the same line, production of 
charmonia via uncorrelated charm quark annihilation in the QGP is expected to
fall significantly below the above computed annihilation yield into gluons, 
lending strong support to our interpretation that all quarkonia are produced
late, when the system reaches the critical temperature and hadronizes.

This result has a number of significant consequences. We first note that the
small annihilation rate in the plasma also implies that charmonia are not
likely to be formed in the plasma. Therefore, the possible existence of
(quasi-)bound $J/\psi$ mesons in the QGP \cite{datta,asa} would not
influence our predictions for charmonia yields as these states would never be
populated during the evolution of the QGP. We neglect in this context $J/\psi$
production in hard collisions before QGP formation, i.e. before $\tau_0$,
because of the the large formation time \cite{huefner}.

These considerations also highlight the differences between our approach,
where all charmonia are formed non-perturbatively at T$_c$, during 
hadronization of the QGP, and the recombination model of \cite{the1,the2}, 
where the cross section for the production of $J/\psi$ mesons from $c$ and 
$\bar{c}$ quarks is obtained by first computing, using the operator product 
expansion technique \cite{peskin}, the  dissociation of $J/\psi$ mesons due 
to collisions with gluons and then inverting this cross section using detailed
balance. We note that this procedure yields a cross section for $J/\psi$ 
production of several mb \cite{the3}, significantly exceeding that for 
annihilation of the $c\bar{c}$ pair into 2 gluons, see Eq.~\ref{e3-1}.   

\section{Centrality and rapidity dependence of quarkonia yields} \label{s_cy}

\begin{figure}[ht]
\hspace{-.7cm}
\begin{tabular}{lr}
\begin{minipage}{.49\textwidth}
\vspace{-.6cm}
\centering\includegraphics[width=1.2\textwidth]{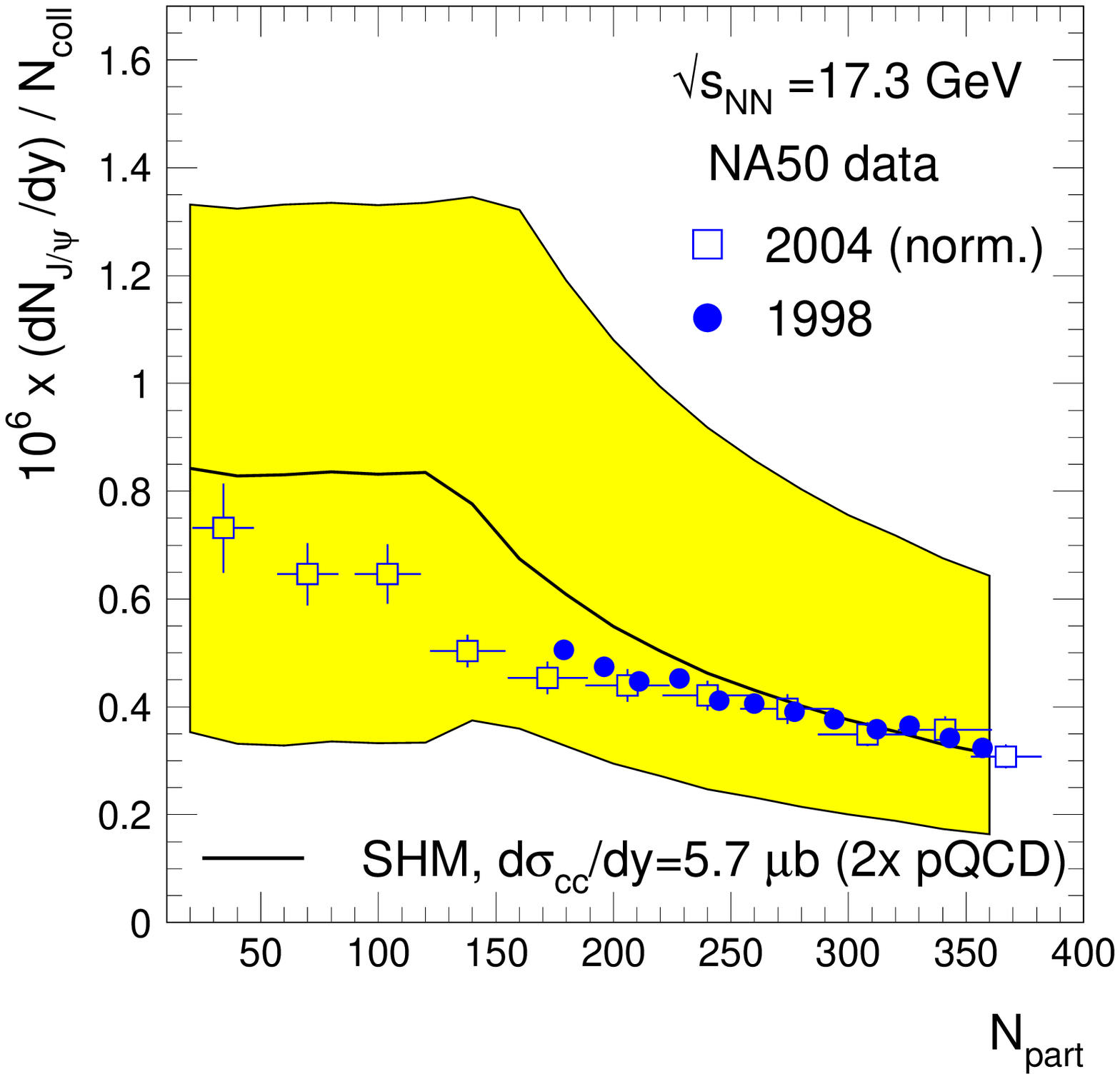}
\end{minipage} \begin{minipage}{.49\textwidth}
\vspace{-.6cm}
\centering\includegraphics[width=1.2\textwidth]{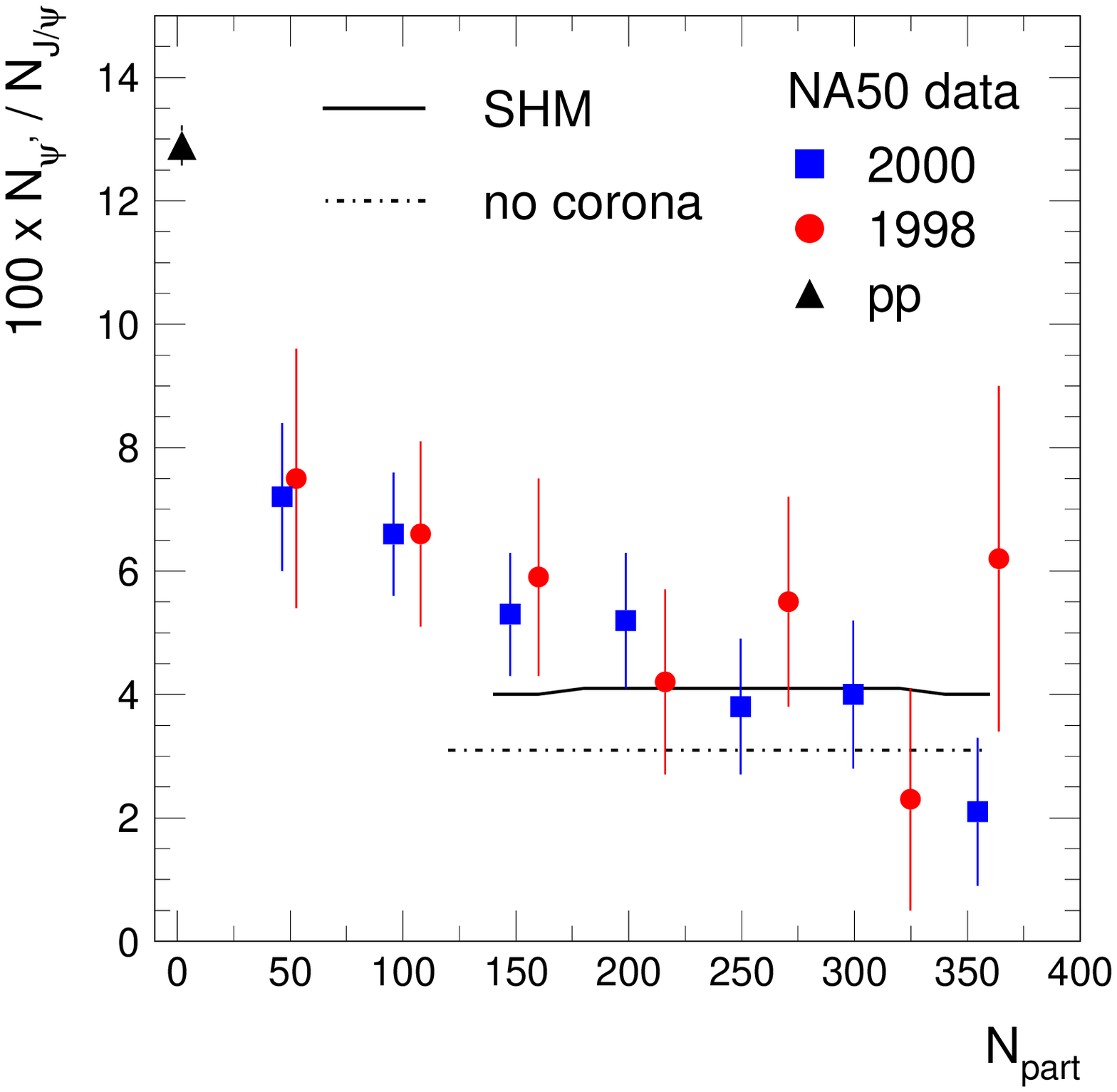}
\end{minipage}
\end{tabular}
\caption{Centrality dependence of $J/\psi$ yield per number of collisions
  (left panel) 
and of the ratio of yields of ${\psi'}$ and ${J/\psi}$ (right panel) at 
the SPS energy. The shaded band in the left panel denotes the range in the 
charm production cross section (a factor of 2 up and down) and the error
assumed for the $J/\psi$ cross section in pp collisions.
The experimental data on $J/\psi$ yield are from ref. \cite{gos} 
(see \cite{pbm1}) and ref. \cite{na50a} (2004 NA50 data) and on the 
${\psi'}/{J/\psi}$ ratio from ref. \cite{na50b,na50c}.}
\label{aa_fig3a}
\end{figure}

In our earlier studies \cite{pbm1,aa1} we have shown that the $J/\psi$ data 
at SPS energy can be described within the statistical approach, but only 
when assuming that the charm production cross section is enhanced by about
a factor of 3 beyond the perturbative QCD (pQCD) predictions. 
Those results were obtained assuming that hadronization of charm takes
place in the whole volume of the system at chemical freezeout.
We consider this assumption to be rather extreme and adopt here a picture
of hadronization within one unit of rapidity at midrapidity, as done
for the analysis of the yields of hadrons without charm   within the thermal
model  \cite{pbm0,aa2}. 
The comparison of our model calculations with the data measured at SPS
by the NA50 experiment is shown in Fig.~\ref{aa_fig3a}.
Two sets of data are included in Fig.~\ref{aa_fig3a}: the NA50 data of 1998,
as analyzed by Gosset et al. ~\cite{gos} and with a further normalization
(see \cite{pbm1}) and the NA50 data of 2004 \cite{na50a}, representing
the yield of $J/\psi$ normalized to the Drell-Yan yield, which we arbitrarily
normalize for the sake of comparison, assuming that the Drell-yan yield
scales with $N_{coll}$. It is seen that the two data sets 
show a very good consistency in the centrality dependence.
Note that a factor of 0.5 was used to convert the full phase space 
experimental data \cite{gos,pbm1}, to midrapidity yields.

The experimental data are well described by invoking a moderate enhancement of
the charm production cross section of a factor of 2 compared to pQCD
calculations \cite{rv1}.  Very clearly, adopting for the present calculations
the hadronization within one unit of rapidity leads to an agreement with the
data for a smaller enhancement of the charm cross section than required in our
earlier study \cite{aa1}.  New measurements by the NA60 experiment
\cite{na60a} indicate that the enhancement in the dimuon yield below the
$J/\psi$ mass, earlier observed by the NA50 \cite{na50en}, is of prompt origin
and, as such, cannot be interpreted as an enhancement of the charm production
cross section, although an experimental result on the charm cross section is
currently not available.  We also note that a factor of 2 enhancement is
within uncertainties of the pQCD calculations.

A comparison between calculations and data \cite{na50b,na50c} is shown in 
the right panel of Fig.~\ref{aa_fig3a} for the centrality dependence of 
the yield of ${\psi'}$ relative to $J/\psi$. Using the present core and 
corona model leads to a 25\% increase of the ratio compared to the case 
when all nucleons are considered to be part of the QGP region (no corona).
The model predictions are shown only down to the $N_{part}$ values
corresponding to $V_{QGP}^{min}$. For more peripheral collisions the data
suggest a continuous transiton towards the value measured in elementary
collisions.

\begin{figure}[ht]
\hspace{-.7cm}
\begin{tabular}{lr}
\begin{minipage}{.49\textwidth}
\vspace{-1cm}
\centering\includegraphics[width=1.2\textwidth]{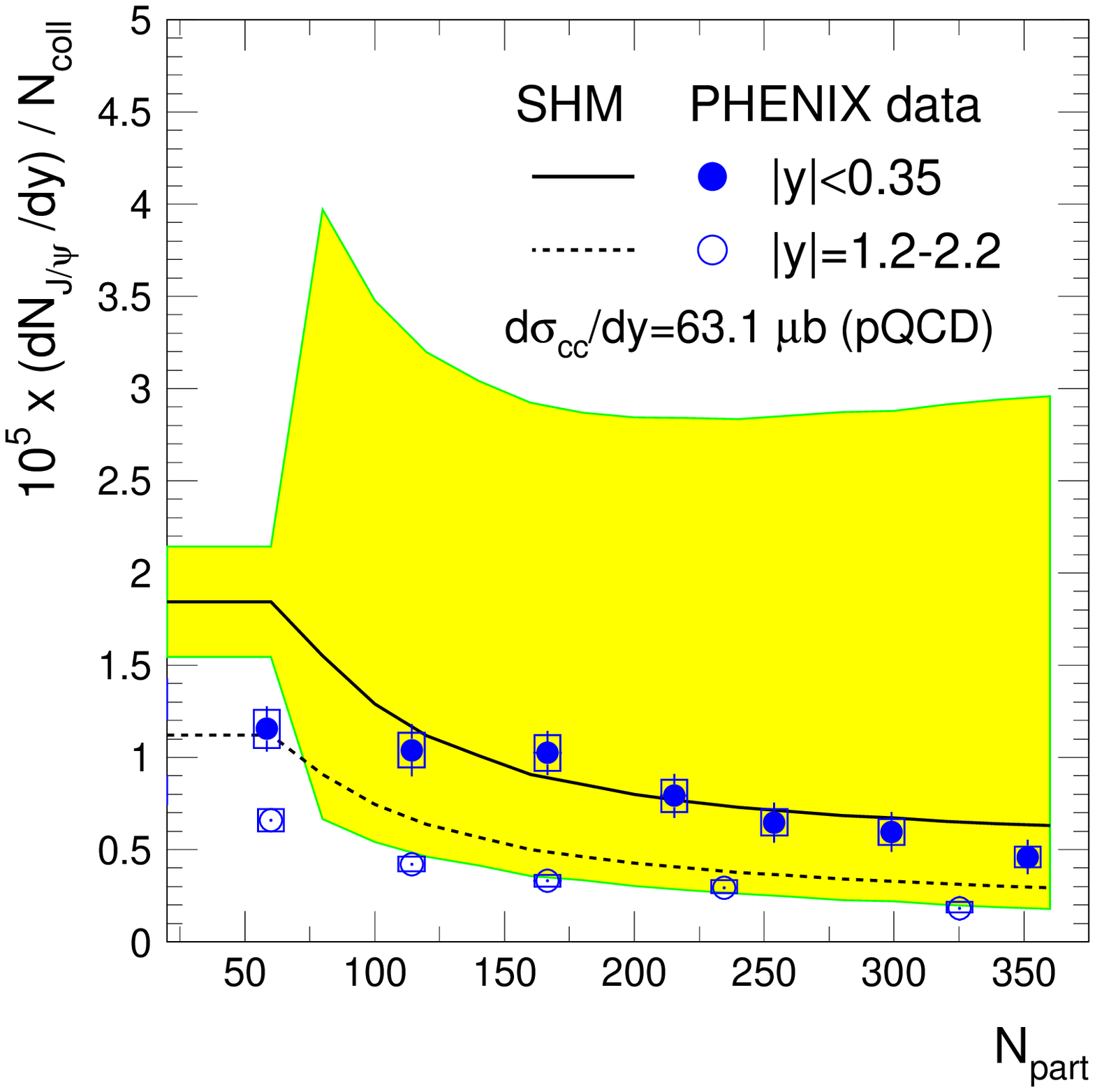}
\end{minipage}  & \begin{minipage}{.49\textwidth}
\vspace{-1cm}
\centering\includegraphics[width=1.2\textwidth]{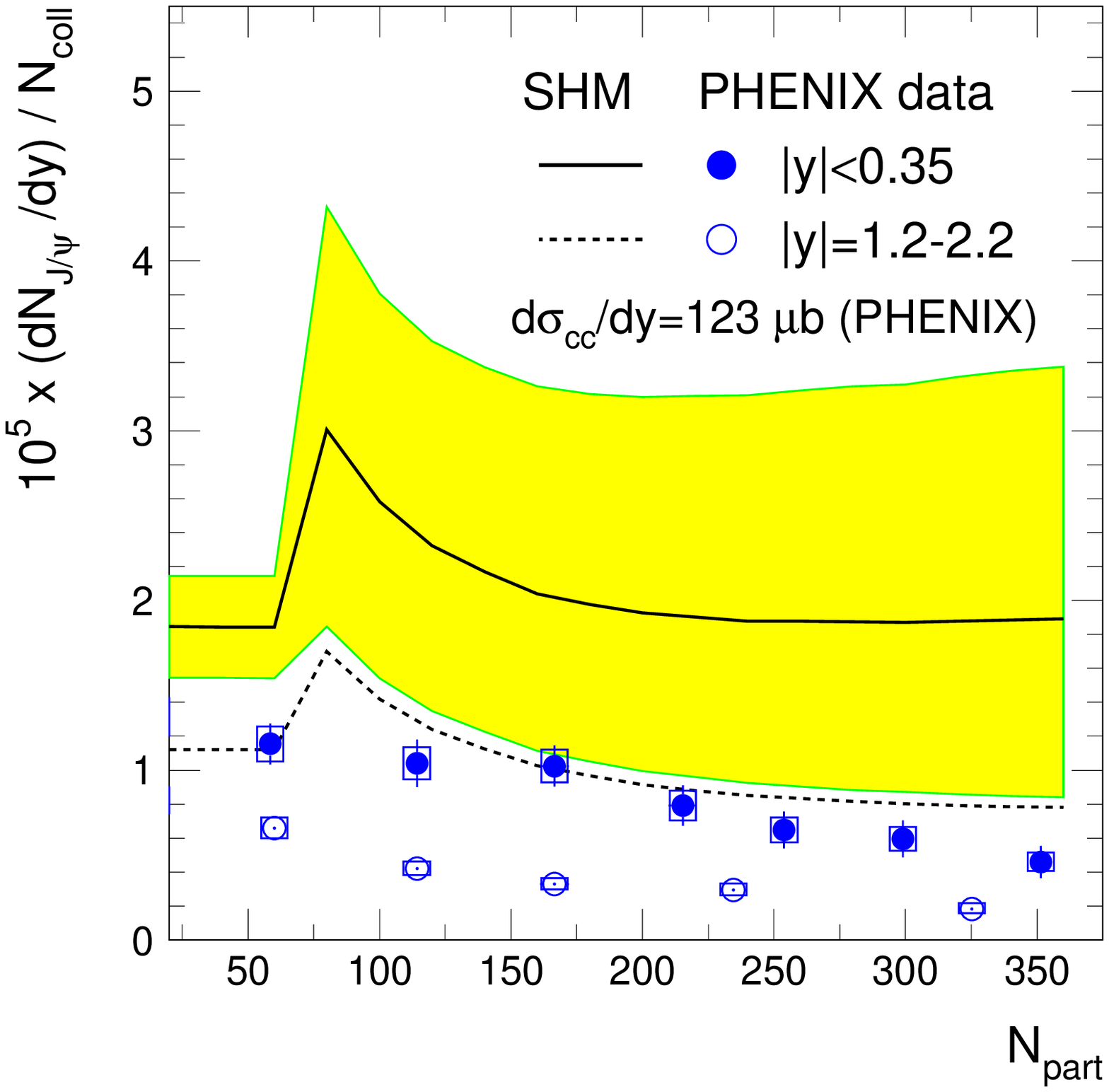}
\end{minipage} \\
\begin{minipage}{.49\textwidth}
\vspace{-1cm}
\centering\includegraphics[width=1.2\textwidth]{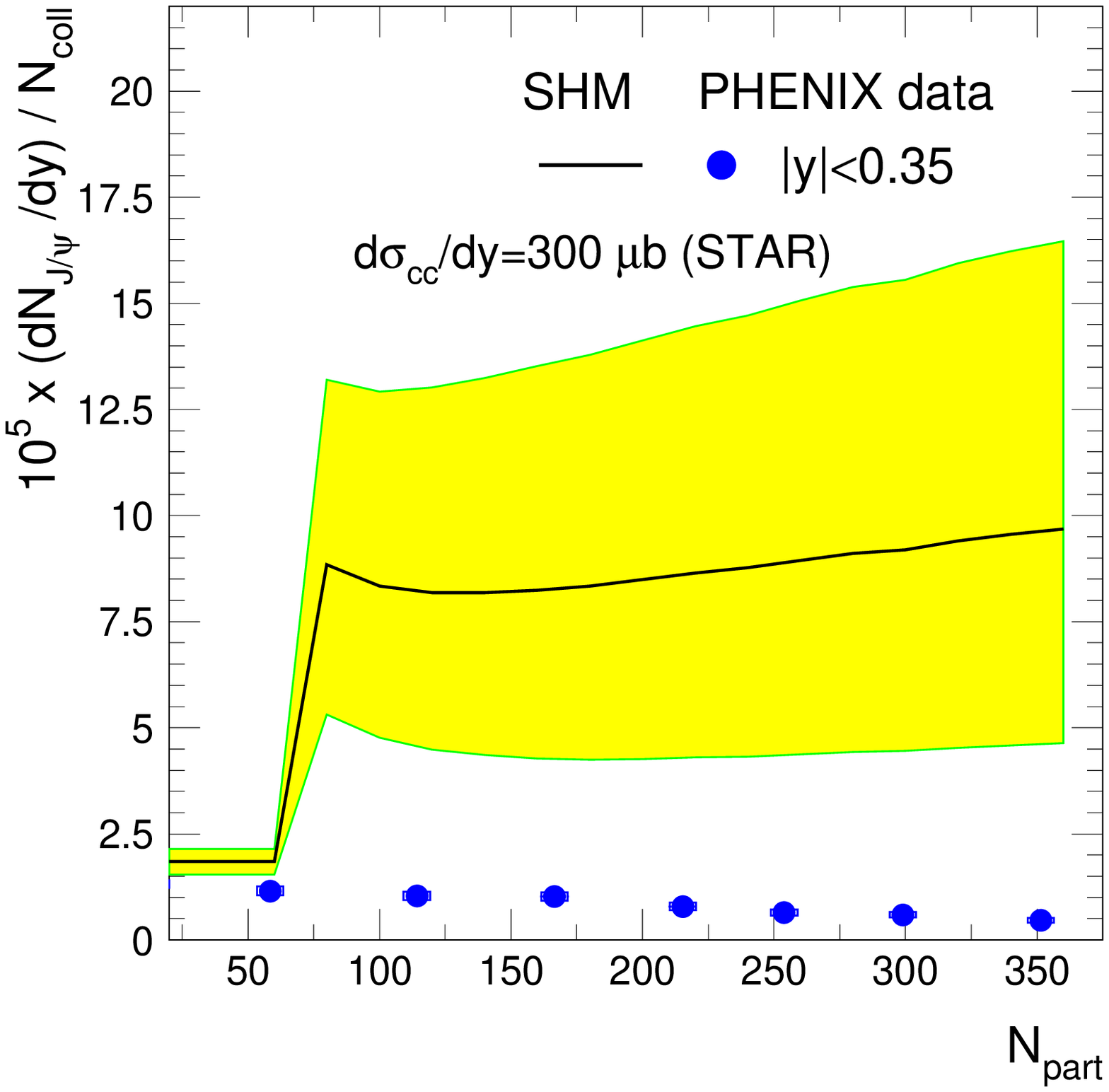}
\end{minipage}  & \begin{minipage}{.44\textwidth}
\vspace{-1cm}
\caption{Centrality dependence of $J/\psi$ yield at midrapidity per number of
  collisions 
at RHIC. 
The experimental data are compared to three cases for the model
calculations, considering the charm production cross section: i) as
calculated in pQCD \cite{cac}, and as measured by ii) PHENIX \cite{cc2} and
iii) STAR \cite{cc1} experiments.  Note the different scales on the vertical
axis. The lines are our calculations for the central value of the charm
production cross section, the band corresponds to its systematic 
uncertainties. The dashed lines are model predictions for the rapidity 
$y$=1.7.}
\label{aa_fig3}
\end{minipage}
\end{tabular}
\end{figure}
%\vspace{-0.2cm}

For  RHIC energy, the centrality dependence of the $J/\psi$ rapidity density
at midrapidity is shown  in Fig.~\ref{aa_fig3}, considering three cases 
for the charm production cross section: 
i) as calculated in pQCD \cite{cac}, and as measured by ii) PHENIX \cite{cc2}
and iii) STAR \cite{cc1}.
The model calculations agree with the recent PHENIX data \cite{per}
very well for the  central value of the pQCD charm production cross section, 
in line with our earlier comparison \cite{aa1}.
The yield is scaled by the number of binary collisions, $N_{coll}$.
In this scaled representation the model predicts a slight decrease 
of the $J/\psi$ yield as a function of $N_{part}$, a trend which is
compatible with the data. 
The model overestimates the data by about a factor of 2 for the central 
value of the charm production cross section measured by PHENIX \cite{cc2}, 
but is compatible with the data for the lower limit of this measured cross 
section.
If we use the values of the charm cross section measured by STAR \cite{cc1} 
the model overestimates the data by about a factor of 10 and is
clearly incompatible with the data also concerning the trend as a function
of centrality.

Note that the steps in the calculations seen around $N_{part}$=70 are the 
outcome of the sharp transition from the core region to the pp region below
$V_{QGP}^{min}$=400 fm$^3$.
As we have mentioned earlier, this transition is probably continuous, 
but we have not tried at this stage to model it.

\begin{figure}[ht]
  \centering\includegraphics[width=.95\textwidth]{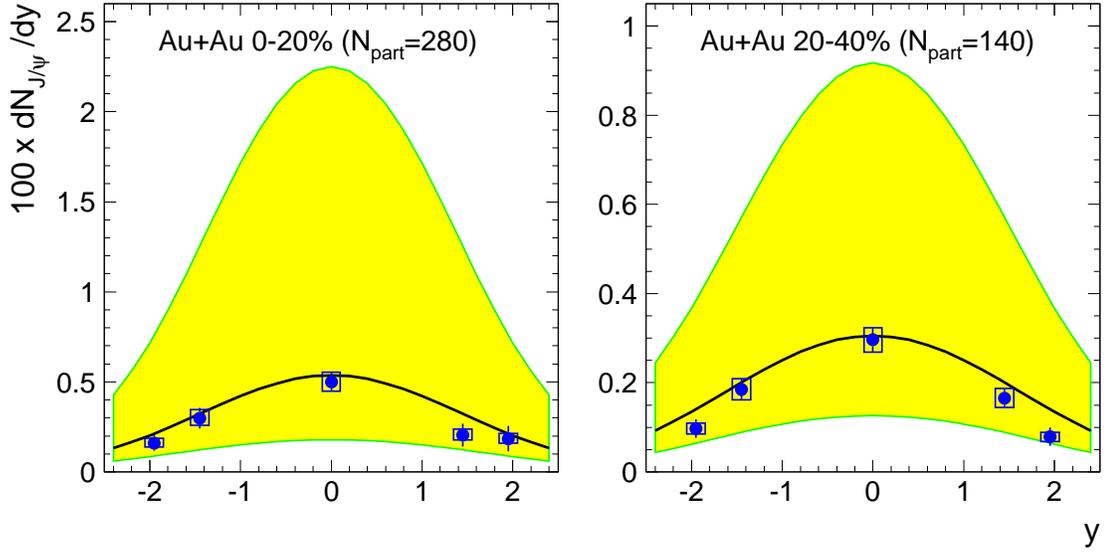}
  \caption{Rapidity dependence of the $J/\psi$ yield at RHIC for two
    centrality  
bins. The SHM calculations, performed for the nominal pQCD charm production
cross section (continuous line with the band denoting the systematic errors
of the cross section) are compared to recent PHENIX data \cite{per}.}
  \label{aa_fig4}
\end{figure}

The rapidity dependence of the $J/\psi$ yield is shown in Fig.~\ref{aa_fig4}
for two centrality bins for Au-Au collisions.
The PHENIX data \cite{per} are well described by the model calculations 
for the central value of the pQCD charm production cross section.
In particular, the narrow rapidity distributions predicted within the
kinetic recombination model \cite{the3} are not observed here because 
of the rather broad rapidity distribution of the charm production cross 
section \cite{cac1}.
Based on the PHOBOS measurement of the pseudorapidity density of charged 
particles \cite{pho2}, we have assumed a constant volume $V_{\Delta y=1}$ 
as a function of rapidity.
As a consequence, the rapidity dependence is determined in our model solely
by the input charm cross section.

\begin{figure}[htb]
\begin{tabular}{lr}
\begin{minipage}{.49\textwidth}
\vspace{-1cm}
\centering\includegraphics[width=1.15\textwidth]{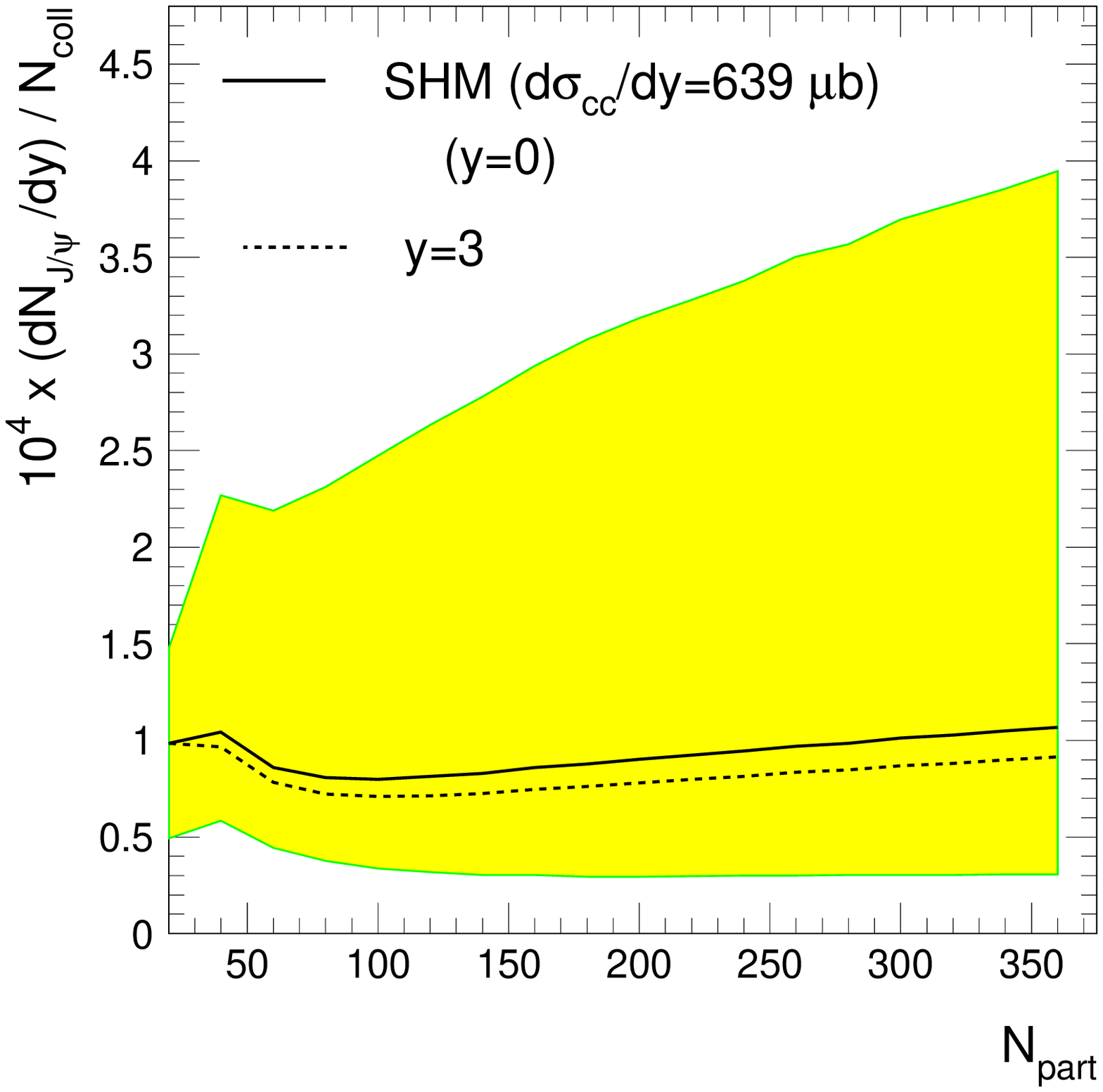}
\end{minipage}  & \begin{minipage}{.49\textwidth}
\vspace{-1cm}
\centering\includegraphics[width=1.15\textwidth]{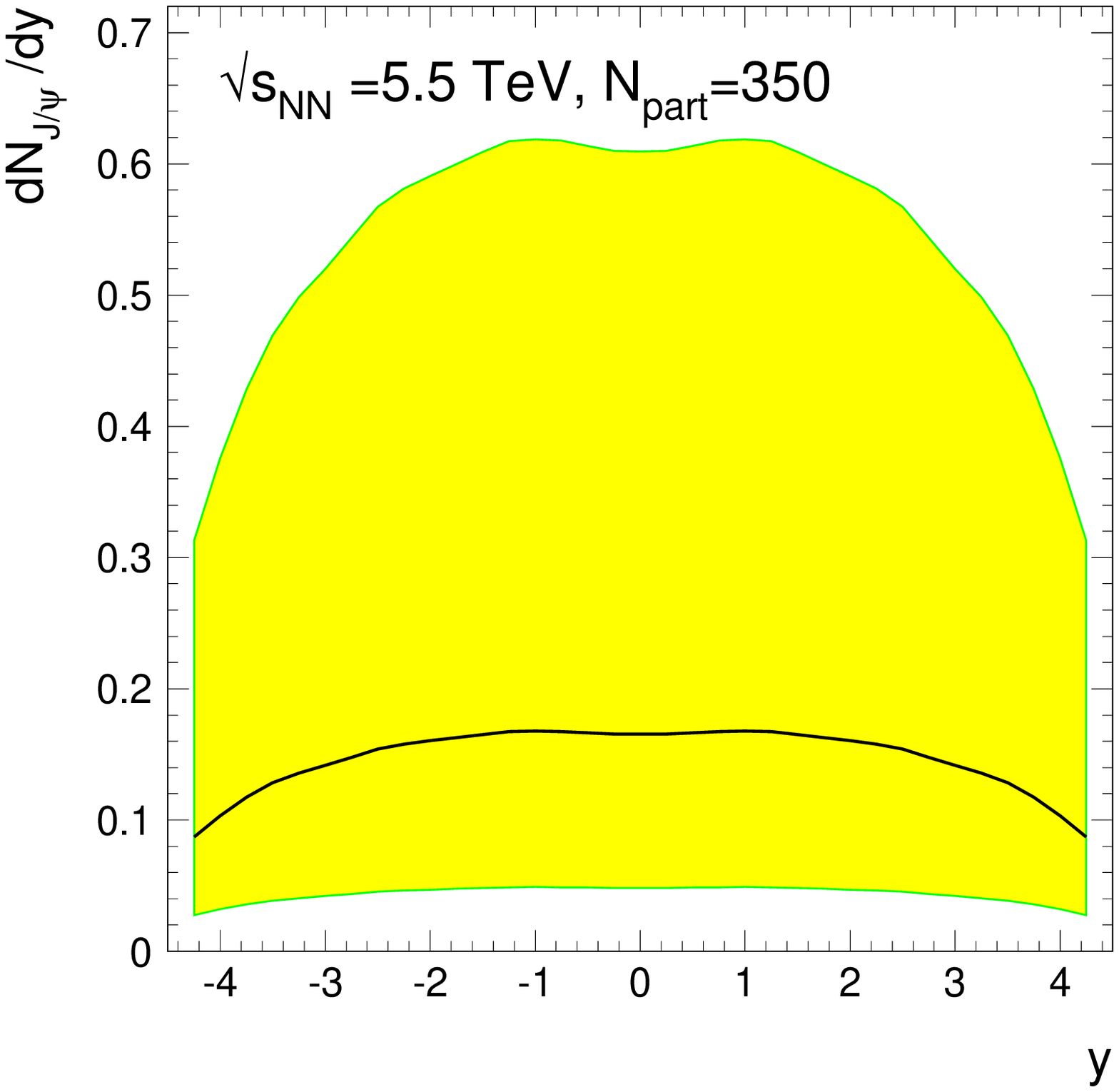}
\end{minipage} \end{tabular}
\caption{Predictions for the $J/\psi$ rapidity density at LHC. 
Left panel: centrality dependence (normalized to $N_{coll}$) at midrapidity,
right panel: rapidity dependence for central collisions ($N_{part}$=350).}
\label{aa_fig5}
\end{figure} 

The model predictions for the centrality and rapidity dependence of $J/\psi$
production at LHC are shown in Fig.~\ref{aa_fig5}. For most of the range of
the charm production cross section considered, a growth with centrality faster
than that of the number of binary collisions is seen is seen in the $J/\psi$
yield, allowing to make a distinct case for the statistical hadronization
scenario at LHC. This increase is stronger for larger charm cross section
values.  The rapidity distribution is broader than at RHIC energy. As a
consequence, the measured rapidity densities of $J/\psi$ in the ALICE central
barrel ($|\eta|<0.9$) and in the muon spectrometer ($\eta=2.5-4.0$) are
expected to be comparable.

\begin{figure}[htb]
\begin{tabular}{lr}
\begin{minipage}{.49\textwidth}
\vspace{-1cm}
\centering\includegraphics[width=1.15\textwidth]{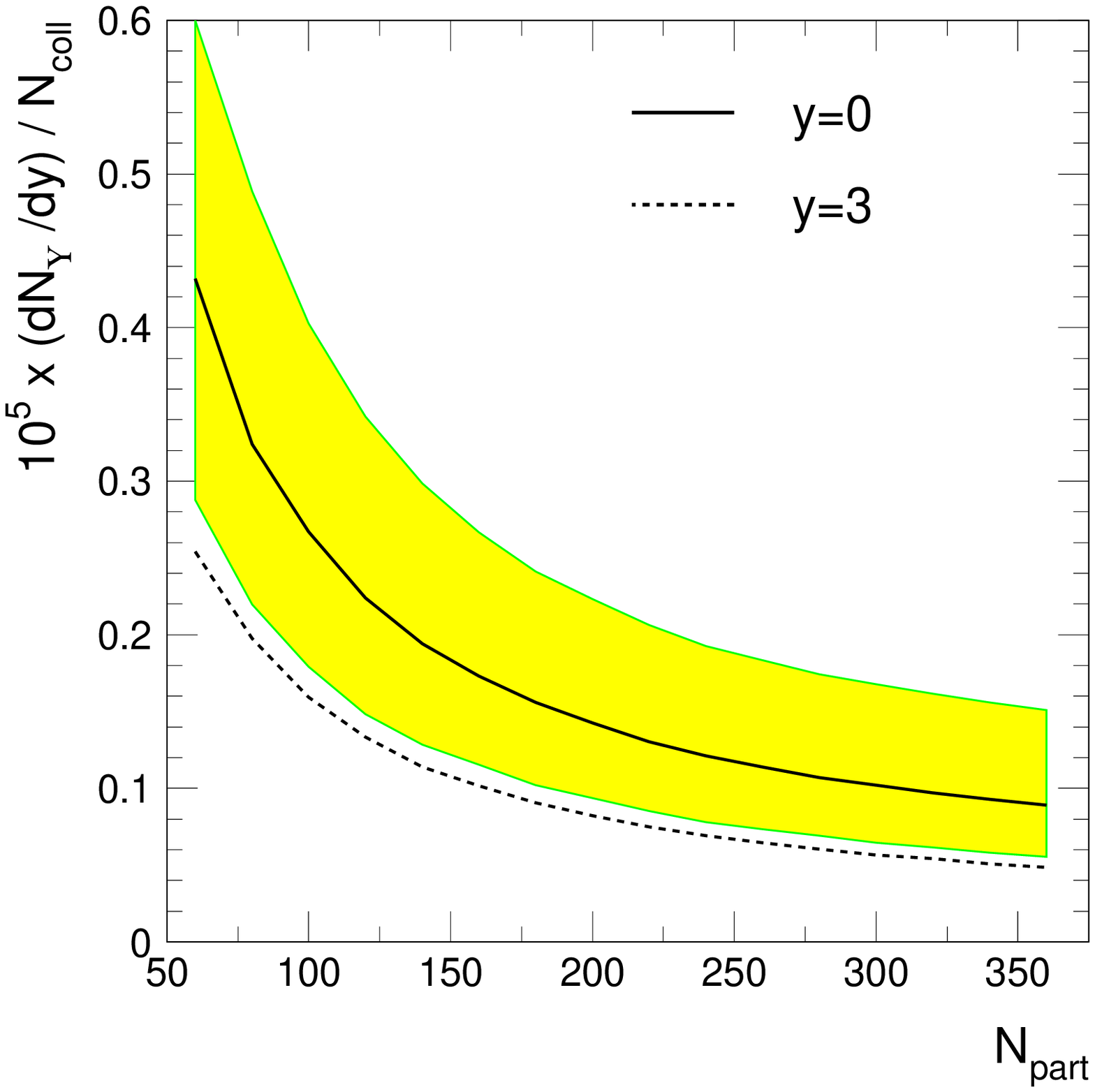}
\end{minipage}  & \begin{minipage}{.49\textwidth}
\vspace{-1cm}
\centering\includegraphics[width=1.15\textwidth]{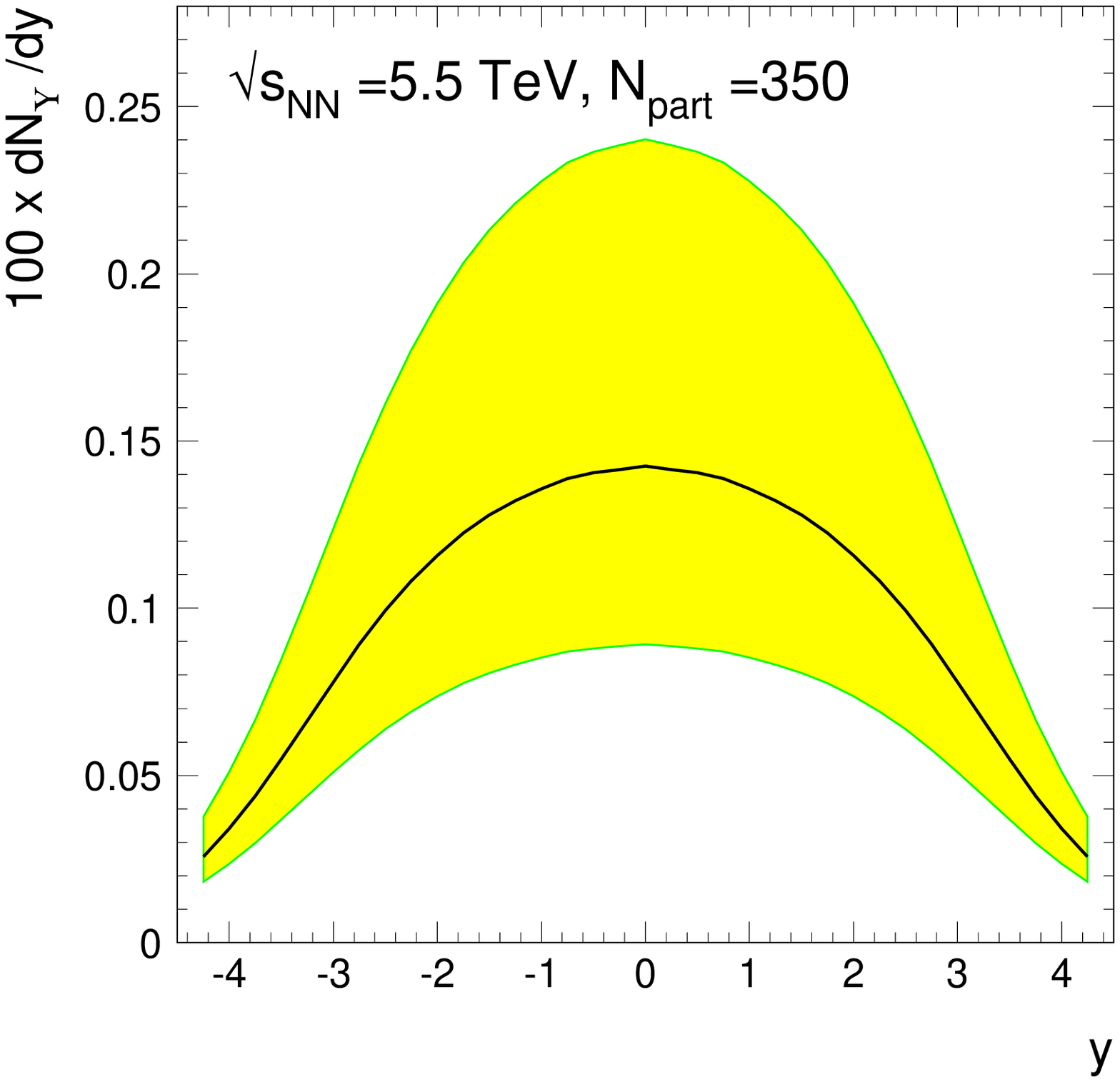}
\end{minipage} \end{tabular}
\caption{Yield of $\Upsilon$ at LHC. Left panel: centrality dependence 
for $y$=0 (continuous line with error band for the uncertainty of 
bottom cross section) and $y$=3 (dashed line). 
Right panel: rapidity dependence for central collisions ($N_{part}$=350). }
\label{aa_fig5x}
\end{figure} 

The model predictions for the centrality and rapidity dependence of 
$\Upsilon$ production at LHC are shown in Fig.~\ref{aa_fig5x}. 
A decrease as a function  of centrality is seen (note again the normalization
to $N_{coll}$), similar to the one observed for $J/\psi$ at RHIC.

\begin{figure}[ht]
\vspace{-1cm}
\centering\includegraphics[width=.7\textwidth]{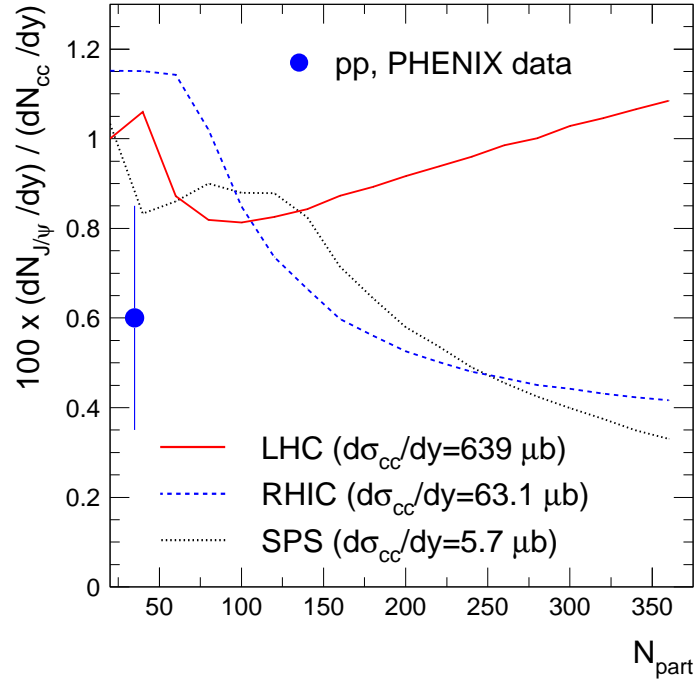}
\caption{Centrality dependence of $J/\psi$ yield normalized to $c\bar{c}$ 
yield for the SPS, RHIC and LHC energies.
The thick line is the value used in the model for pp collisions.
The dot corresponds to the PHENIX measurements in pp collisions at RHIC
energy.}
\label{aa_fig6}
\end{figure}

Fig.~\ref{aa_fig6} shows a comparison of the calculated $J/\psi$ yield per
number of $c\bar{c}$ pairs for RHIC and LHC energies. This figure illustrates
the succinct change in trend expected for measurements at LHC energy, due to
the strong increase with energy of the total charm production cross section.
Note, in particular the decrease with centrality of the $J/\psi$ yield at
RHIC, originating from 
the canonical suppression of the open charm hadrons.  The dot corresponds to
the PHENIX measurement of $J/\psi$ production in pp collisions \cite{phe3}
normalized to the
$c\bar{c}$ cross section, also measured by PHENIX \cite{cc2}. This
value is smaller than the value of 1\% derived using the pQCD value \cite{cac}
for the charm cross section and much smaller than the value of 2.5\% derived
earlier by Gavai et al. \cite{gav}.

\begin{figure}[!hbt]
\begin{tabular}{lr}\begin{minipage}{.49\textwidth}
\vspace{-.5cm}
\centering\includegraphics[width=1.17\textwidth]{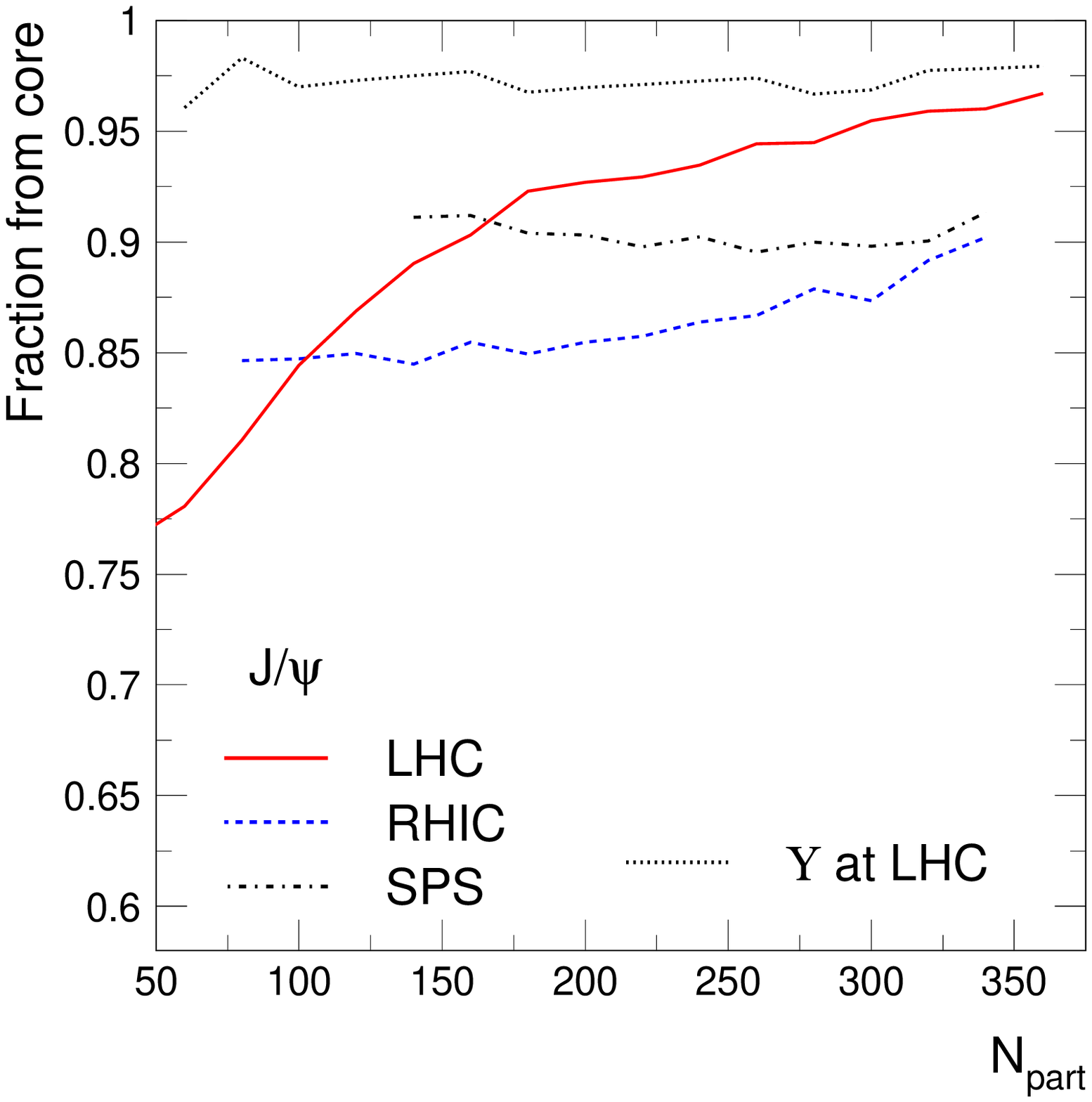}
\end{minipage}  & \begin{minipage}{.49\textwidth}
\vspace{-.5cm}
\centering\includegraphics[width=1.17\textwidth]{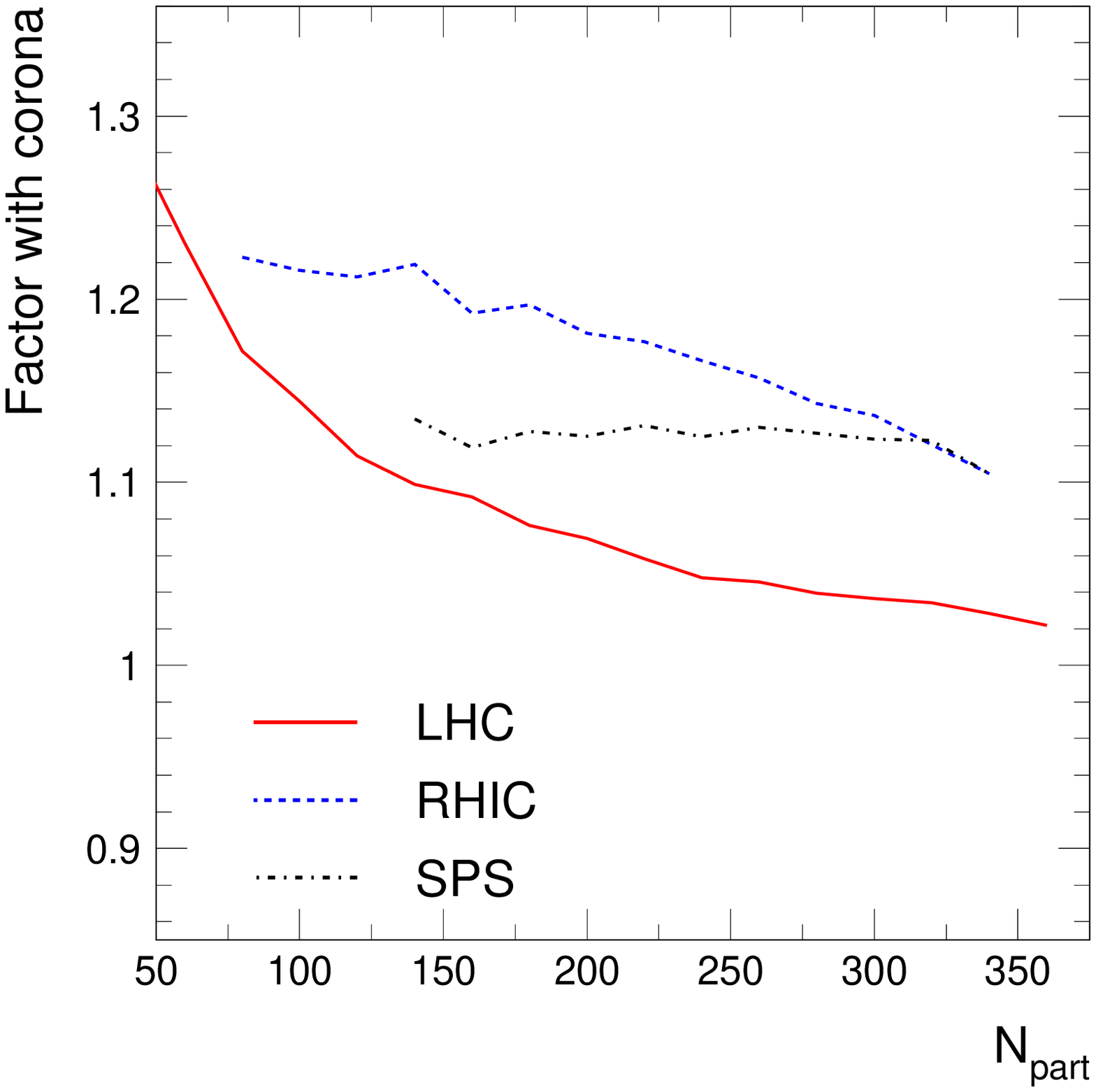}
\end{minipage} \end{tabular}
\caption{Centrality dependence of corona contribution.
Left panel: the fraction of yields of $J/\psi$ and $\Upsilon$ from the core
(QGP) region relative to the total overlap.
Right panel: the increase of the $J/\psi$ yield considering corona, compared
to the case when all the overlap volume is QGP.}
\label{aa_fig6x}
\end{figure}

In the left panel of Fig.~\ref{aa_fig6x} we present the fraction of $J/\psi$
and of $\Upsilon$ yields from the core (QGP) region relative to the total 
(core and corona) overlap. For the $J/\psi$ yields, this contribution of 
corona to the total $J/\psi$ yield is moderate for all the three energies. 
Note as well the different centrality dependences for the three energy
regimes. The $\Upsilon$ yield is predominantly originating from the core
part.
The right panel of Fig.~\ref{aa_fig6x} shows the factor by which the $J/\psi$ 
yield increases when considering the corona contribution, compared to the 
case when we assume that the entire overlap volume is QGP.
The curves reflect the differences between the yields calculated in our model
and the yields in pp collisions (see Fig.~\ref{aa_fig6}). 
Since we consider a rather modest corona fraction in our calculations, the
increase of the $J/\psi$ yield due to the corona is moderate.

\section{Transverse momentum dependence} \label{s_pt}

We turn now to the analysis of the transverse momentum ($p_t$) spectrum of
$J/\psi$ mesons. Within our model, the $p_t$ distribution of $J/\psi$, as well
as of any hadron carrying charm (or beauty), is determined by the temperature
and transverse expansion velocity, $\beta$, at chemical freeze-out.  We employ
a formula for collective expansion \cite{sch}, but consider, to keep the input
minimal, one average velocity instead of a velocity profile:

\begin{equation}
\frac{\ud N}{p_t \cdot \ud p_t}\sim m_t \cdot 
I_0\biggl(\frac{p_t\mathrm{sinh}y_t}{T} \biggl) \cdot 
K_1\biggl(\frac{p_t\mathrm{cosh}y_t}{T} \biggl)
\label{eq:pt}
\end{equation}
where $m_t=\sqrt{m_0^2+p_t^2}$, $m_0$ is the rest mass of the particle,
and $y_t=\mathrm{tanh}^{-1}(\beta)$, with $\beta$ the collective (transverse)
expansion velocity in units of the speed of light. 
The $I_0$ and $K_1$ are the modified Bessel functions.

\begin{figure}[hbt]
\centering\includegraphics[width=.62\textwidth]{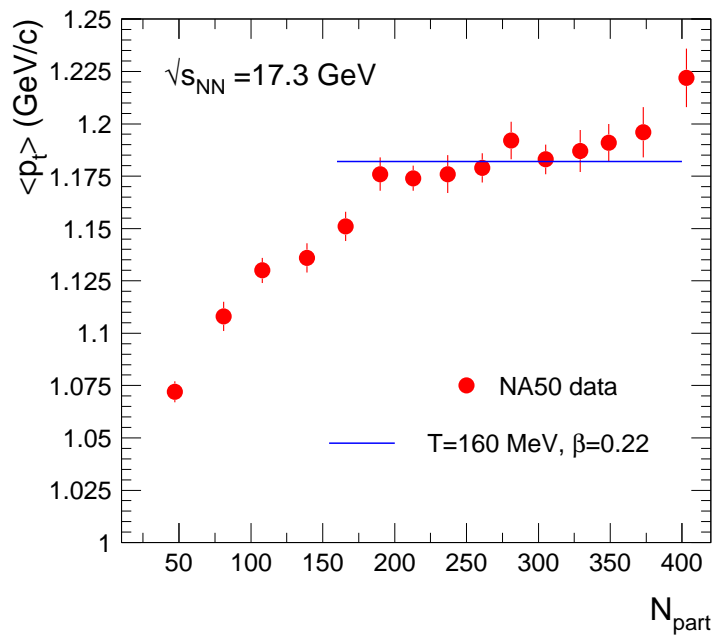}
\caption{Centrality dependence of the average transverse momentum of 
$J/\psi$ mesons in Pb-Pb collisions at SPS energy. 
The data measured by the NA50 experiment \cite{na50pt} are compared to 
calculations using Eq.~\ref{eq:pt}.}
\label{aa_fig7a}
\end{figure} 

Our picture is similar to that tested with data at the SPS by Gorenstein 
et al. \cite{bug1} (first proposed by Grandchamp and Rapp \cite{gra1}). 
Spectra of both $J/\psi$ and $\psi'$ mesons and of $\Omega$ hyperons were 
shown \cite{bug1} to be described by a (kinetic) temperature equal to 
the chemical freeze-out temperature, while the expansion velocity was found
to be substantially lower than for the other hadrons.
This indeed suggests a kinetic freeze-out of $J/\psi$ coincident with the 
chemical freeze-out (at the phase boundary between QGP and hadronic
matter \cite{aa2}), with $\beta$ characterizing the transverse expansion in
the QGP phase. 

To check the validity of this interpretation, we calculate with 
Eq.~\ref{eq:pt} the transverse momentum spectrum of $J/\psi$ and compare
it to the available data.
For SPS energy we show in Fig.~\ref{aa_fig7a} the centrality dependence of 
the average transverse momentum of $J/\psi$ measured by the NA50 experiment
\cite{na50pt}. Lacking a measured $J/\psi$ spectrum in pp collisions,
we have not included the corona contribution in our calculations.
The data are well described by the calculations using $T$=160 MeV and 
$\beta$=0.22, in agreement with the earlier results \cite{bug1}. 
Note that the data approach a constant value for central collisions
($N_{part}>$150), where the present model is applicable.
We would like to emphasize that the $\langle p_t\rangle$
data cannot constrain $T$ and $\beta$ in a satisfactory way. 
When fitting the data, local minima are present in the $\chi^2$ 
distribution. Several sets of $T$ and $\beta$ values describe the data
equally well, as known from fits of distributions of other hadron species
\cite{star}.
This problem would be alleviated if high-precision spectra as a function
of centrality become available.
We note that the centrality dependence of the average transverse momentum 
measured at SPS \cite{na50pt}, Fig.~\ref{aa_fig7a}, is well explained over
the whole centrality range assuming a purely hadronic scenario
\cite{gavin,kha2,arm} (see discussion in ref. \cite{cro2}).

\begin{figure}[hbt]
\vspace{-.4cm}
\centering\includegraphics[width=.7\textwidth]{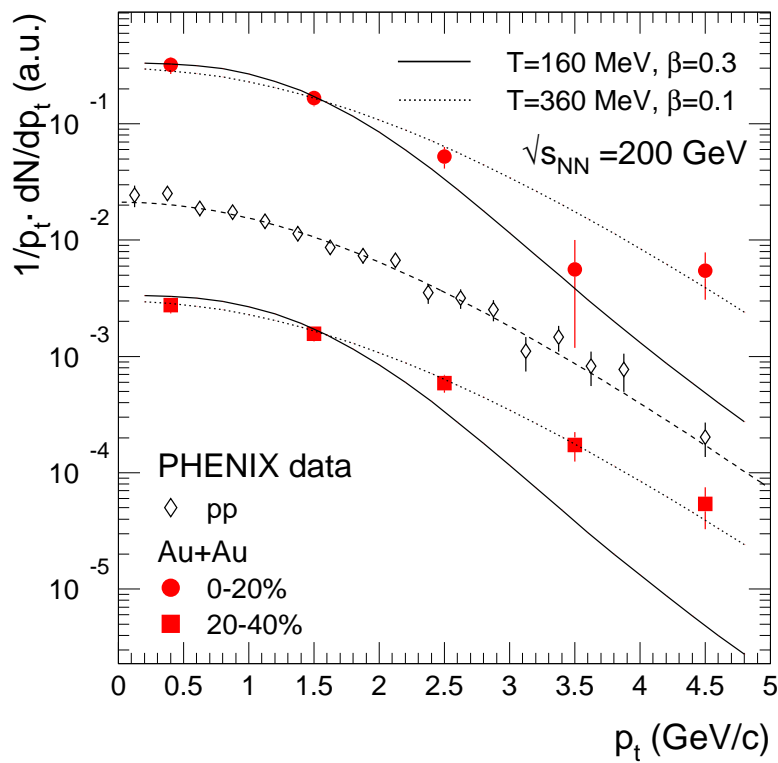}
\caption{Transverse momentum distribution of $J/\psi$ in Au-Au collisions 
at  RHIC energy \cite{per}. The  data (symbols) are compared 
with calculations using Eq.~\ref{eq:pt} for two cases of the kinetic
freezeout parameters (lines). The spectrum measured in pp collisions 
\cite{phe4} with the fit as used for the corona part in our calculations
is also shown.}
\label{aa_fig7}
\end{figure} 

For  RHIC energy we calculate the $J/\psi$ spectrum taking into account
both the QGP contribution (according to Eq.~\ref{eq:pt}) and the corona part.
For the corona, we use the spectrum measured in pp collisions \cite{phe4}, 
which we fit in order to be able to interpolate or extrapolate.
Fig.~\ref{aa_fig7} we show the comparison of the $J/\psi$ spectra measured 
by the PHENIX experiment \cite{per} to our calculations.
Two sets of parameters are investigated for the QGP part: 
i) $T$=160 MeV, $\beta$=0.3 and
ii) $T$=360 MeV, $\beta$=0.1.
The data are compatible with the picture of $J/\psi$ production
via hadronization at chemical freeze-out, expected to be described by 
scenario i).
In detail, the data appear to exhibit a high-pt component which is not 
accounted for by the corona component used in our model and which seems
better described by a larger temperature in the QGP region.
Interestingly, the expansion velocity extracted from fits of the spectra of
$\Omega$ and $\phi$ particles, which are also expected to interact rather 
weakly in the hadronic stage, yields for $T$=160 MeV $\beta\simeq$0.4-0.5,
significantly lower than $\beta$ values determined for other hadrons 
\cite{star}.
We note that spectra of open charm mesons, as inferred from the measured
nuclear modification factor of non-photonic electrons \cite{v2b2} are
also compatible with a thermal distribution at chemical freeze-out with a
flow velocity $\beta\simeq$0.3 as used here for the $J/\psi$ spectra.

\begin{figure}[htb]
\begin{tabular}{lr} \begin{minipage}{.49\textwidth}
\centering\includegraphics[width=1.16\textwidth]{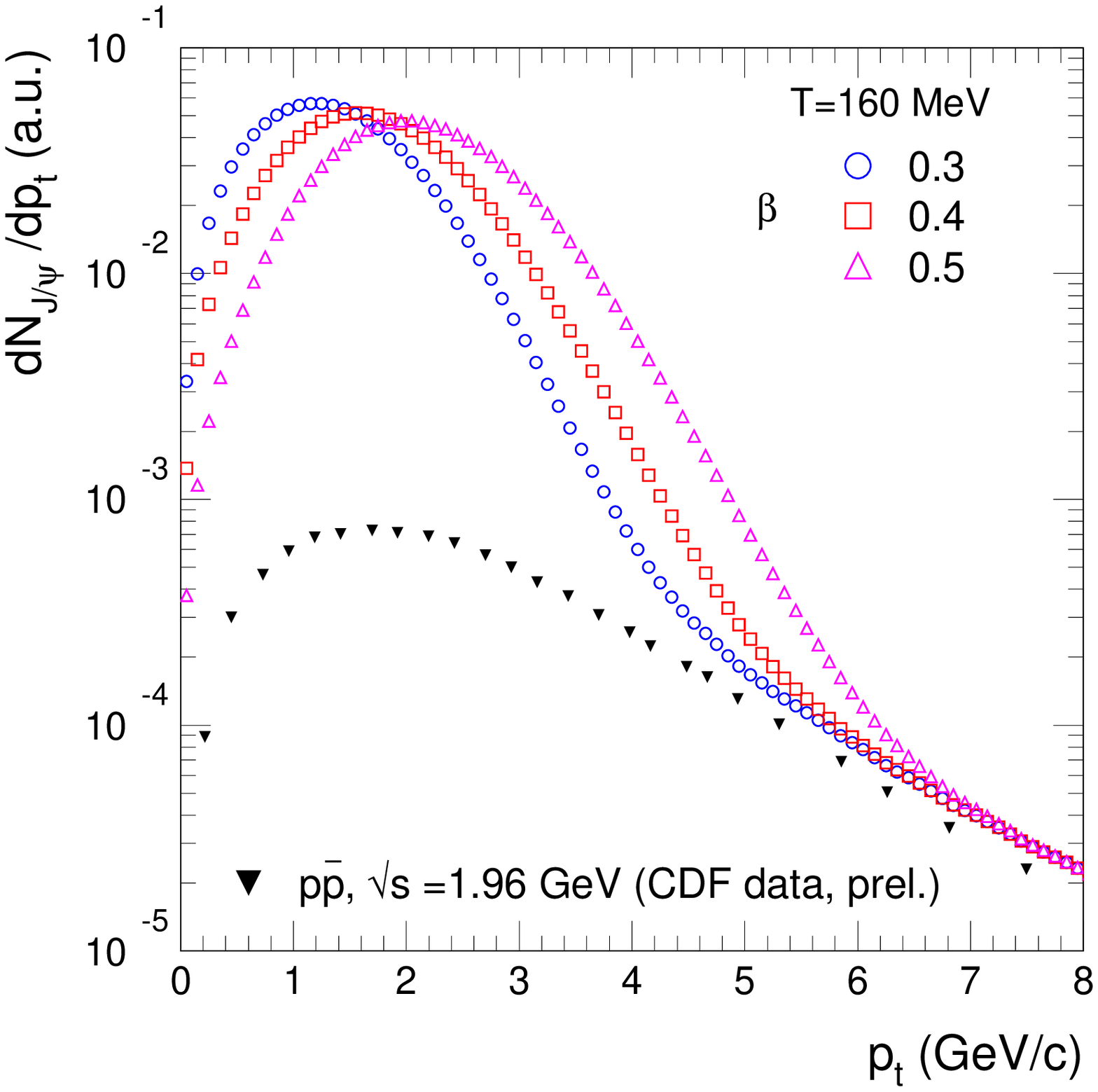}
\end{minipage}  & \begin{minipage}{.49\textwidth}
\centering\includegraphics[width=1.16\textwidth]{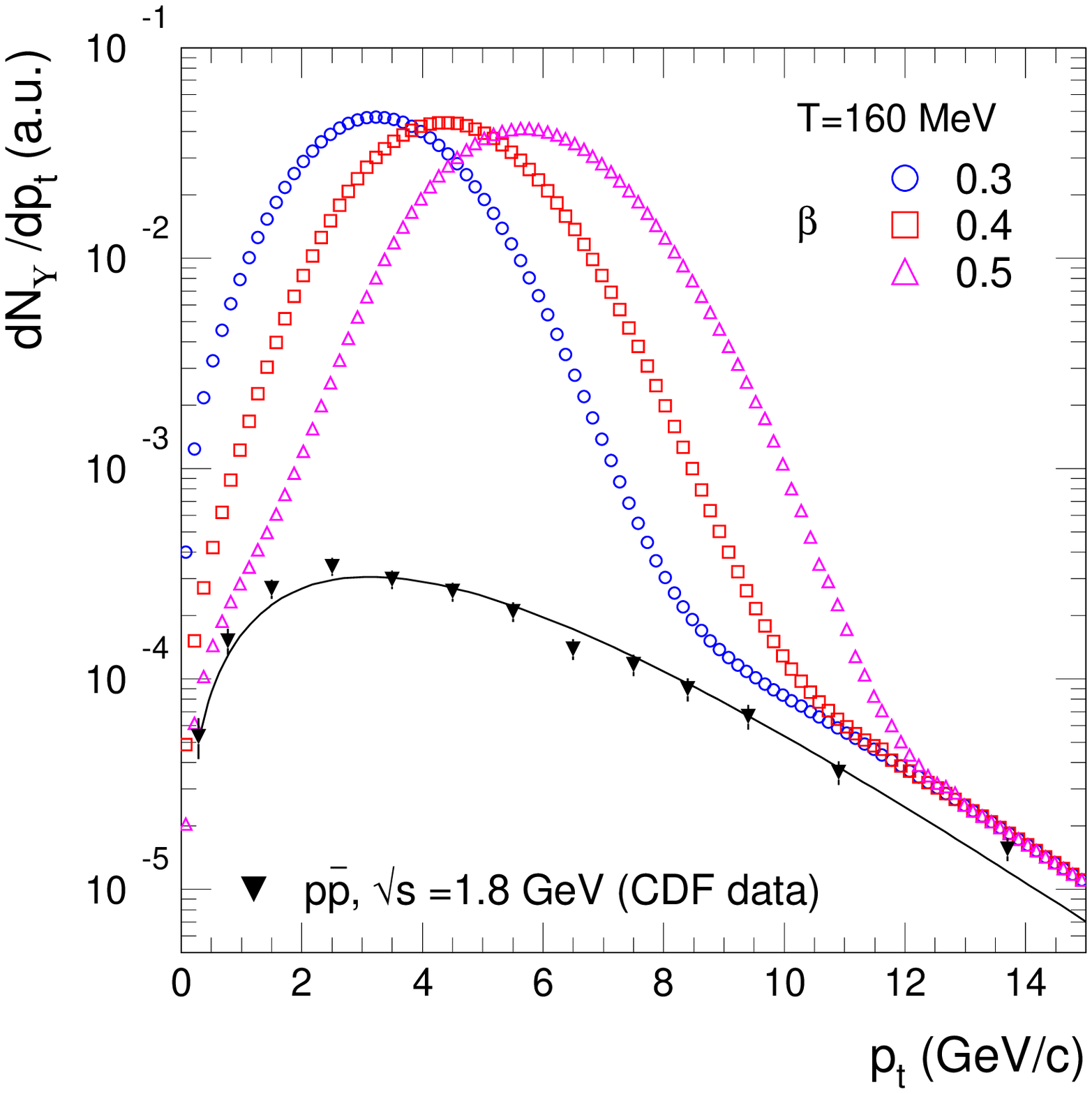}
\end{minipage} \end{tabular}
\caption{Predictions for momentum spectrum of $J/\psi$ (left panel) 
and $\Upsilon$ (right panel) for different values of the expansion velocity, 
$\beta$, for central Pb+Pb collisions ($N_{part}$=350). Also included are
the measured spectra in p$\bar{\mathrm{p}}$ collisions at Tevatron \cite{pau}.}
\label{aa_fig7x}
\end{figure}

In Fig.~\ref{aa_fig7x} we present predictions for the transverse momentum 
spectrum of $J/\psi$ and $\Upsilon$ in central collisions at LHC. 
We show the spectra for a range of expansion velocities, as expected  
for LHC energy.
For pp collisions in the corona we have assumed the same spectral shape 
at LHC as measured in p$\bar{\mathrm{p}}$ collisions at Tevatron 
\cite{pau}, which are also included for comparison in Fig.~\ref{aa_fig7x}.
The corona fractions corresponding to central Pb+Pb collisions ($N_{part}$=350)
are 0.13 for $J/\psi$ and 0.09 for $\Upsilon$.
For both $J/\psi$ and $\Upsilon$, the spectral shapes expected within the 
statistical hadronization model are very different compared those 
expected from the superposition of pp collisions, making the data at LHC 
a stringent test of our model.

\section{Conclusions}

We have presented new results on the statistical hadronization of heavy quarks 
at the SPS, RHIC and LHC energies.
We have focused on the results for $J/\psi$ yields, but have also extended 
the model to predict $\Upsilon$ yields in Pb+Pb collisions at the LHC energy.
Included in our model is now a separation of the collision geometry into 
a core (QGP) and a corona (pp collisions) part.
Our estimate of the annihilation rates of charm quark in a hot plasma
demonstrates that this is a small effect.
An important ingredient in our model is the charm production cross section,
which is presently quite poorly measured. We have considered the effect of
its uncertainty on the model predictions.
For  SPS energy, our present calculations are performed assuming that 
the hadronization takes place within one unit of rapidity at midrapidity, 
as used also for  RHIC and LHC energies. 
Good agreement with data is found with a charm cross section which is 
moderately (a factor of 2) larger, but consistent within errors with pQCD 
calculations.
For  RHIC energy we have studied the centrality and rapidity dependence
of $J/\psi$ production. Our model is in good agreement with the available
data, lending strong support for the picture of statistical hadronization. 
We have discussed the transverse momentum distributions of $J/\psi$ mesons 
expected from the model. The present data, in particular those at RHIC energy,
are compatible with the picture of the $J/\psi$  
(kinematic) freeze-out coincident with  chemical freeze-out, but more 
precise data are needed to quantitatively explore this idea.
Our predictions for  LHC energy concerning centrality, rapidity and 
transverse momentum dependence of the $J/\psi$ yield will be tested with 
data within the next few years.

\section*{Acknowledgments}
We are indebted to M. Cacciari for providing us the rapidity distribution 
of the charm production cross section at the RHIC energy and for valuable
discussions.

K. Redlich acknowledges the support from the Polish Committee of Scientific 
Research KBN under grant 2P03 (06925).

\end{document}